\documentclass[sigconf,authorversion]{acmart}

\usepackage[caption=false]{subfig}
\usepackage{url}
\usepackage{threeparttable}
\usepackage{ragged2e}
\usepackage{float}
\usepackage{xfrac}
\usepackage[frozencache,cachedir=.]{minted}
\usepackage{censor}

\usepackage{pdflscape}
\usepackage{multirow}
\usepackage{multicol}

\setminted {
    frame=lines,
    framesep=2mm
}

\def\numPolices{28}
\def\numDatasets{3}

\def\numTotalPwds{205,176,321}

\def\graphSizeMultiplier{0.75}
\def\figureSizeMultiplier{0.75}
\def\codeSampleFontSize{\normalsize}
\def\tableFontSize{\footnotesize}

\def\PCP{\ensuremath{\phi}}
\def\supp#1{\mathsf{supp}(#1)}
\def\suppPCP#1#2{\mathsf{supp}_{#1}(#2)}
\def\maxPCP#1#2{\mathsf{max}_{#1}(#2)}
\def\surplus#1#2{\mathsf{surplus}(#1, #2)}
\def\fresh#1#2#3#4{\mathsf{fresh}(#1, #2, #3, #4)}
\def\magnitude#1{\mathsf{mag}(#1)}

\def\reselection#1#2#3{\mathsf{Reselection}(#1, #2, \textsf{#3})}
\def\Skeptic{\textsc{Skeptic}}
\def\Authority{\textsc{Authority}}
\def\Pyrrho{\textsc{Pyrrho}}
\def\Pacpal{\textsc{Pacpal}}
\def\Passlab{\textsc{Passlab}}

\makeatletter
\def\mdseries@tt{m}
\makeatother

\AtBeginDocument{%
  \providecommand\BibTeX{{%
    \normalfont B\kern-0.5em{\scshape i\kern-0.25em b}\kern-0.8em\TeX}}}

\copyrightyear{2020}
\acmYear{2020}
\setcopyright{acmlicensed}\acmConference[ASIA CCS '20]{Proceedings of the 15th ACM Asia Conference on Computer and Communications Security}{October 5--9, 2020}{Taipei, Taiwan}
\acmBooktitle{Proceedings of the 15th ACM Asia Conference on Computer and Communications Security (ASIA CCS '20), October 5--9, 2020, Taipei, Taiwan}
\acmPrice{15.00}
\acmDOI{10.1145/3320269.3384762}
\acmISBN{978-1-4503-6750-9/20/10}

\begin{document}
\fancyhead{}

\title{Skeptic: Automatic, Justified and Privacy-Preserving Password Composition Policy Selection}

\author{Saul Johnson}
\email{saul.johnson@tees.ac.uk}
\orcid{0000-0001-9876-3775}
\affiliation{%
  \institution{Data Analytics Research Group \\
  Teesside University}
  \city{Middlesbrough}
  \country{UK}
}
\author{Jo\~ao F. Ferreira}
\email{joao@joaoff.com}
\orcid{0000-0002-6612-9013}
\affiliation{%
  \institution{INESC-ID and Instituto Superior T\'ecnico \\
  University of Lisbon}
  \city{Lisbon}
  \country{Portugal}
}
\author{Alexandra Mendes}
\email{alexandra@archimendes.com}
\orcid{0000-0001-8060-592}
\affiliation{%
  \institution{HASLab, INESC TEC, Porto and \\Department of Informatics \\ 
  Universidade da Beira Interior}
  \city{Covilh\~a}
  \country{Portugal}
}
\author{Julien Cordry}
\email{j.cordry@tees.ac.uk}
\orcid{0000-0002-6489-3026}
\affiliation{%
  \institution{Immersive Tech. Research Group \\
  Teesside University}
  \city{Middlesbrough}
  \country{UK}
}


\renewcommand{\shortauthors}{S. Johnson, J. F. Ferreira, A. Mendes, and J. Cordry}
\begin{abstract}
The choice of password composition policy to enforce on a password-protected system represents a critical security decision, and has been shown to significantly affect the vulnerability of user-chosen passwords to guessing attacks. In practice, however, this choice is not usually rigorous or justifiable, with a tendency for system administrators to choose password composition policies based on intuition alone. In this work, we propose a novel methodology that draws on password probability distributions constructed from large sets of real-world password data which have been filtered according to various password composition policies. Password probabilities are then redistributed to simulate different user password reselection behaviours in order to automatically determine the password composition policy that will induce the distribution of user-chosen passwords with the greatest uniformity, a metric which we show to be a useful proxy to measure overall resistance to password guessing attacks. Further, we show that by fitting power-law equations to the password probability distributions we generate, we can justify our choice of password composition policy without any direct access to user password data. Finally, we present \Skeptic{}---a software toolkit that implements this methodology, including a DSL to enable system administrators with no background in password security to compare and rank password composition policies without resorting to expensive and time-consuming user studies. Drawing on \numTotalPwds{}~passwords across \numDatasets{}~datasets, we lend validity to our approach by demonstrating that the results we obtain align closely with findings from a previous empirical study into password composition policy effectiveness.
\end{abstract}




\maketitle

\section{Introduction}\label{sec:introduction}
If we wish to increase the resilience of a password-protected system to password guessing attacks, our thoughts might turn in the first instance to password composition policies---sets of rules that dictate which subset of the space of all supported passwords users are permitted to create on the system. The selection of a suitable password composition policy, however, has not historically been carried out according to rigorous selection criteria \cite{nist2006electronic}, with a tendency for system administrators to base their decision on which one ``feels'' like it would lead to a more secure distribution of user-chosen passwords. For such a critical component of system security, the finding that the restrictiveness of a password composition policy has little to no correlation with the value of the assets it protects is somewhat alarming \cite{florencio2010where}, and makes a strong case for a more rigorous method of selection.

While much study to date has been conducted on how password composition policies affect the security of password-protected systems, such work usually consists of an analysis of either leaked data sets that have since been released into the public arena \cite{weir2010testing} or of passwords that have been collected under different password composition policies specifically for the purpose of the study \cite{komanduri2011passwords,shay2016designing}. The former condition means that it is very difficult to estimate how some of the more exotic password composition policies affect system security because databases of passwords created under those policies are not available. While it might be tempting to merely filter these datasets according to the policy we wish to examine, previous work \cite{kelley2012guess} finds that this does not create a dataset that is representative of one that is actually created under that policy, with passwords in filtered datasets tending to be stronger. The latter condition, while allowing password security researchers to collect data under any password composition policy they choose, has considerably less ecological validity; the participants were, after all, creating passwords in an experimental setting and not on any real-world system of value to them as individuals. Gathering and analysing data in this way is also expensive, time-consuming, and requires significant domain expertise, placing it beyond the reach of a typical system administrator working in the field. Finally, both of these methodologies raise privacy concerns. In each case, we are handling user-generated passwords that may still be in use by those individuals, or else be usable to infer passwords that are. We are motivated, therefore, to search for a methodology that permits us to automatically choose a suitable password composition in a way that allows us to justify that choice while avoiding the propagation of the user password data that informs it. This is especially important considering the recent rise in previously-leaked passwords being employed in phishing scams \cite{schofield2019phishing} against the users they belong to.

In this work, we propose such a methodology, and present \Skeptic{}---a software toolchain that puts it into practice. We begin by drawing on large sets of leaked password data \cite{cubrilovic2009rockyou,gross2012yahoo,burgess2016check} to derive password probability distributions. By redistributing password probabilities in different ways, we can simulate different modes of password reselection behaviour that might be exhibited by users when forced to select a different password by the password composition policy. Drawing on work by Malone and Maher \cite{malone2012investigating} and Wang et al. \cite{wang2017zipfs}, we fit power-law curves to these password probability distributions, allowing us to quantify the additional guessing attack resistance conferred by their associated password composition policies in isolation from the password data itself. Following related research into increasing system security by maximising password diversity \cite{segreti2017diversity,malone2012investigating,blocki2013optimizing}, we achieve this by using the \textit{uniformity} of these distributions as a proxy for their overall resistance to guessing attacks. To maximise the practical utility of the data we generate, \Skeptic{} includes the Password Composition Policy Assertion Language (\Pacpal{})---a DSL for straightforwardly comparing and ranking password composition policies using this data.

Using a selection of password composition policies drawn from related work and data from three large-scale password data breaches, we demonstrate our methodology and its implementation (as the \Skeptic{} toolchain) by rigorously and justifiably ranking password composition policies under a range of different assumptions about user password reselection behaviour. As our evaluation data, we use \numDatasets~datasets containing a total of \numTotalPwds~passwords, studying \numPolices~distinct password composition policies. The results we obtain correlate strongly with those from previous empirical studies on the effects of password composition policies on the security of user-chosen passwords, with some interesting findings that warrant further study. For instance, we find that stricter (i.e. less usable) password composition policies dramatically reduce password probability distribution uniformity if we assume that user password reselection behaviour will converge on a small number of remaining permitted passwords. We further demonstrate that the \Skeptic{} toolchain supports straightforward specification 
of password composition policies 
from within the \textit{Coq} proof assistant, with all the advantages we would expect from such an encoding, including the ability to check from within \textit{Coq} that certain password composition policies confer immunity to the \textit{Mirai} and \textit{Conficker} botnet malware.

We have introduced the work and its motivation in this Section~\ref{sec:introduction}. In Section~\ref{sec:background}, we introduce related work. We then move on to describing our methodology in detail in Section~\ref{sec:methodology}, in a manner designed to facilitate implementation to encourage replication and experimentation. In Section~\ref{sec:skeptic} we describe the implementation of our methodology as the \Skeptic{} toolchain. Section~\ref{sec:evaluation} contains an evaluation of our approach, in which we attempt to replicate previous empirical research \cite{shay2016designing} on password composition policy effectiveness. Finally, we conclude in Section~\ref{sec:conclusion}.

\section{Related Work}\label{sec:background}
There exists a wealth of password data online that has been compromised from various sources and released into the public arena. \citeauthor{weir2010testing}~\cite{weir2010testing} draw on a few different sets of this data in order to examine the validity of using password entropy as defined in NIST document SP800-63-1 \cite{nist2006electronic} to determine the security provided by various password composition policies. The authors conclude, based on experiments run against some of the same datasets we use in this work \cite{cubrilovich2009rockyou}, that it is not a valid metric, empirically validating earlier work by Verheul \cite{verheul2006selecting} proving that conversion of Shannon entropy-like measures into password guessing entropy under different password composition policies is not possible. Such work demonstrates the effective use of large breached datasets in password composition policy research, and in Section~\ref{subsec:replic-weir} we replicate a subset of its results as part of validation of our novel methodology.

It is also possible to use these breached user credential databases straight away to inform our choice of password policy by simply prohibiting all the passwords we can find in them outright. The \textit{Pwned Passwords} web application and API \cite{hunt2018pwned} provides this functionality as a service, aggregating over 500 million unique passwords that have been exposed in data breaches and made publicly available online. Just because a password has not been exposed before, however, does not mean that it is a good password. At the time of writing, for instance, \textit{``breakfast321''} is not present in \textit{Pwned Passwords} but as a dictionary word and run of sequential digits is very likely to be cracked with minimal effort by any of the great number of password cracking algorithms in widespread use today \cite{weir2009password,xu2017password}, with the popular \textit{zxcvbn} password strength checking library \cite{wheeler2016zxcvbn} estimating that this particular example could be cracked in around $10^5$ guesses---well within the capabilities of even the lowliest attacker. The inadequacy of blacklist-based measures alone motivates work such as ours, which aims to equip system administrators with tooling to evaluate the security of arbitrary rule-based password composition policies.

Other studies such as that by \citeauthor{shay2016designing}~\cite{shay2016designing} actively recruit users to create passwords under various password composition policies, and attempt to quantify the security of those policies by running password cracking attacks against passwords collected under these policies. This is considered by many to represent the gold standard of password composition policy research, and as such we replicate results from \citeauthor{shay2016designing}~\cite{shay2016designing} in Section~\ref{subsec:replic-shay} to validate our novel methodology.

Regardless of how it is obtained, it is of vital importance that any model designed to evaluate the effectiveness of password composition policies in reducing the vulnerability of human-chosen passwords to guessing attacks is in some way informed by human-generated password data. Password choice varies significantly across different user demographics (age and nationality for example \cite{bonneau2012science}) and by extension across password-protected systems which have user bases comprising different proportions of these demographics. By consequence of this variability, there can be no definitive password composition policy that will lead to ideal security or usability outcomes across all systems---such policies must be designed on a system-by-system basis. Work by Galbally et al. \cite{galbally2017new} reaffirms this idea---no password strength estimation metric is ideal for all passwords under all conditions. With this in mind, the methodology presented in our work is designed to be attack-independent, and provide a general idea of the security of password composition policies when deployed ``in the wild'' where the shape of password guessing attacks the system might be subjected to can seldom be known in detail. The only assumption we make about the threat model we face is that the attacker is attempting to guess more common passwords first.

As the weakest passwords are, ostensibly, those that are the most likely to be chosen by users, we can think of the ideal password composition policy as the one that induces the most uniform password distribution on our system. Password policies with poor usability will cause users to converge on fewer easy-to-remember passwords and those with poor security will permit the selection of very weak passwords such as ``password'' and ``123456''. This is not a novel argument. Work on adaptive password composition policies \cite{segreti2017diversity} supports the view that greater password diversity is key to system security while research into  password composition policy optimisation \cite{blocki2013optimizing} focuses on maximising minimum password entropy---that is, reducing the probability of the most likely password, analogous to increasing password distribution uniformity. Malone and Maher \cite{malone2012investigating} highlight that user-chosen password distributions are non-uniform, and mention that if this were not the case, attacks that rely on attempting to guess common passwords would become less effective.


\section{Methodology}\label{sec:methodology}
In this section, we present our methodology for rigorous and justifiable password composition policy selection in detail, beginning with raw password data and ending with arbitrary user-specified password composition policies ranked under various assumptions about user behaviour.

\subsection{Sourcing Human-Chosen Passwords}\label{subsec:sourcing-human-chosen-passwords}
With the variability of user password choice in mind \cite{bonneau2012science}, our methodology is parametric on an \textit{input set}---some collection of password data that we expect to be representative of the user base we are modelling, given as a password frequency distribution. Input sets can be sourced from any user credential database where the password plaintext is known, but those used within this work include:

\begin{itemize}
    \item \textbf{RockYou}---compromised in plaintext from the \textit{RockYou} online gaming service of the same name around the year 2009 \cite{cubrilovich2009rockyou}. The password composition policy in place at the time enforced a minimum length of 5 characters with no other requirements \cite{golla2018accuracy}. The version we obtained contained 32,603,048 passwords.
    \item \textbf{Yahoo}---compromised in plaintext from the Yahoo Voices online publishing platform around the year 2012 \cite{gross2012yahoo}. The password composition policy in place at the time of the breach enforced a minimum length of 6 characters with no other requirements \cite{mayer2017second}. The version we obtained contained 453,492 passwords.
    \item \textbf{LinkedIn}---compromised from the professional social networking site \textit{LinkedIn} around the year 2012. The true extent of this breach was unknown until 2016 when it was revealed to be much more extensive than was initially made public \cite{burgess2016check}. Unsalted password hashes in SHA-1 format were compromised, and $\approx98\%$ of these have subsequently been cracked. It is these cracked passwords that make up the LinkedIn dataset we use in this work. The policy in place at the time of the breach enforced a minimum length of 6 characters \cite{mayer2017second}. The version we obtained contained 172,428,238 passwords.
\end{itemize}

\subsubsection{Data cleansing} For a dataset to be as representative as possible, each password within it must have been created by a human under a known password composition policy which has a permitted password space that is a superset of that of the password composition policies we wish to model. It is therefore useful to filter these datasets according to the password composition policy they were created under in order to remove any passwords created under old password composition policies or non-password artifacts \cite{kelley2012guess} that might be present within them. In cases where this policy is not known, it is possible to attempt to infer it using a password composition policy inference tool such as \textit{pol-infer}~\cite{johnson2019inference}. Each dataset was first filtered according to the password composition policy it is known to have been created under. The small proportion of passwords containing non-ASCII characters were then removed to avoid encoding issues that might arise due to multi-byte characters being stored as multiple characters, artificially inflating password length. Some passwords in the Yahoo dataset (10,654 passwords) appeared to be single sign-on flags for integration with an external service, and were accordingly removed. Likewise, some passwords in the LinkedIn dataset (174,088 passwords) appeared to be hexadecimal data (perhaps due to encoding issues), and were also removed. The sizes of each dataset used in this study after this filtration step are shown in Table~\ref{tbl:compliance-data}.

\begin{table}[ht] \tableFontSize
    \centering
    \caption{A breakdown of the number of passwords filtered from each dataset used in this study.}
    \label{tbl:compliance-data}
    \begin{tabular}{|l|l|l|}
        \hline
        Dataset                                  & Filtered size  & Removed           \\ \hline
        RockYou \cite{cubrilovich2009rockyou}    & 32,506,433     & 96,615 (0.30\%)    \\ \hline
        Yahoo \cite{gross2012yahoo}              & 434,287        & 19,205 (4.23\%)   \\ \hline
        LinkedIn \cite{burgess2016check}         & 172,235,601      & 192,637 (0.11\%)    \\
        \hline
    \end{tabular}
\end{table}

\subsubsection{Frequencies to probabilities} 
Following Blocki et al.~\cite{blocki2018economics},
given a cleansed input set $I$ of $N$ user passwords, we
use $f_i$ to denote the frequency of the $i^{th}$ most common password in the set and $pwd_i$ to denote the $i^{th}$ most common password in the set.

The set $I$ induces a probability distribution $D$ over passwords defined as:
\[
D(p) = 
\begin{cases}
\dfrac{f_i}{N} & \text{if } p=pwd_i \\
0              & \text{otherwise}
\end{cases}
\]
The probability $D(p)$ is the probability that a random user selects password $p$. 
We define the magnitude of the distribution induced by $I$ as the number of passwords in $I$. That is, $\magnitude{D} = N$.

\subsection{Specifying Password Composition Policies} \label{subsec:specifying-pcps}
Our methodology is not tied to any specific representation of password composition policies. 
Similar to Blocki et al. \cite{blocki2013optimizing}, we use a set-theoretic notation, with $p \in \phi$ indicating that a password $p$ is permitted by a password composition policy $\phi$. Later on in Section~\ref{subsec:authority}, when we describe our encoding of password composition policies in \Skeptic{}, we will demonstrate that this affords us the power to encode password composition policies for arbitrary software, and scaffold code for doing so automatically.

\subsubsection{Policies studied in this work} \label{subsubsec:pols-studied} We selected and modelled a selection of password composition policies based on those by Shay et al.~\cite{shay2016designing} and Weir et al.~\cite{weir2010testing}, and follow the naming convention used by Shay et al.~\cite{shay2016designing} as follows:

\begin{itemize}
    \item \textbf{basic7, basic8, basic9, basic12, basic14, basic16, basic20}: to comply with policy \textit{basicN}, password must be $N$ characters or greater in length. No other requirements.
    \item \textbf{digit7, digit8, digit9, digit10}: to comply with policy \textit{digitN}, password must be $N$ characters or greater in length, and contain at least one numeric digit.
    \item \textbf{upper7, upper8, upper9, upper10}: to comply with policy \textit{upperN}, password must be $N$ characters or greater in length, and contain at least one uppercase letter.
    \item \textbf{symbol7, symbol8, symbol9, symbol10}: to comply with policy \textit{symbolN}, password must be $N$ characters or greater in length, and contain at least one non-alphanumeric character.
    \item \textbf{2word12, 2word16}: to comply with policy \textit{MwordN}, password must be $N$ characters or greater in length and consist of at least $M$ strings of one or more letters separated by a non-letter sequence.
    \item \textbf{2class12, 2class16, 3class12, 3class16}: to comply with policy \textit{NclassM}, password must be $M$ characters or greater in length and contain at least $N$ of the four character classes (uppercase letters, lowercase letters, digits and symbols).
    \item \textbf{dictionary8}: to comply with policy \textit{dictionaryN} password must be $N$ characters or greater in length. When all non-alphabetic characters are removed the resulting word cannot appear in a dictionary, ignoring case (we used the Openwall ``tiny'' English wordlist \cite{openwall2011wordlists}).
    \item \textbf{comp8}: to comply with policy \textit{compN} password must comply with \textit{dictionaryN} and additionally must contain uppercase letters, lowercase letters, digits and symbols. Replicates the NIST comprehensive password composition policy \cite{nist2013electronic}.
\end{itemize}

\subsection{Modelling Password Reselection} \label{subsec:modelling-pwd-resel}
If a potential user is forbidden from selecting their preferred password by the password composition policy, they must select a different, compliant password or find themselves unable to use the service at all. In this way, a password composition policy induces a change in the probability distribution of passwords on the system. 

In this section, we consider the change induced in a probability distribution $D$ by imposing a password composition policy \PCP. In what follows, we write $\supp{D}$ to denote the support of distribution $D$, that is:
\[
\supp{D} = \{\,p \,|\, D(p) \ge 0\,\}
\]
and we write $\suppPCP{\PCP}{D}$ to denote the support of $D$ restricted to passwords that comply with \PCP:
\[
\suppPCP{\PCP}{D} = \{\,p \,|\, p \in \supp{D} \wedge p \in \PCP\,\}
\]
We assume that $\suppPCP{\PCP}{D}$ will always be non-empty.

The change induced in $D$ by \PCP\ can be seen as a redistribution of the probabilities associated with passwords that do not comply with the password composition policy. 
The sum of the probabilities that need to be redistributed is denoted as $\surplus{D}{\PCP}$ and defined as:
\[
\surplus{D}{\PCP} = \sum_{\substack{p \in \supp{D} \\ p \not\in \PCP}}{D(p)}
\]

\begin{figure}[ht]
    \includegraphics[width=\graphSizeMultiplier\columnwidth]{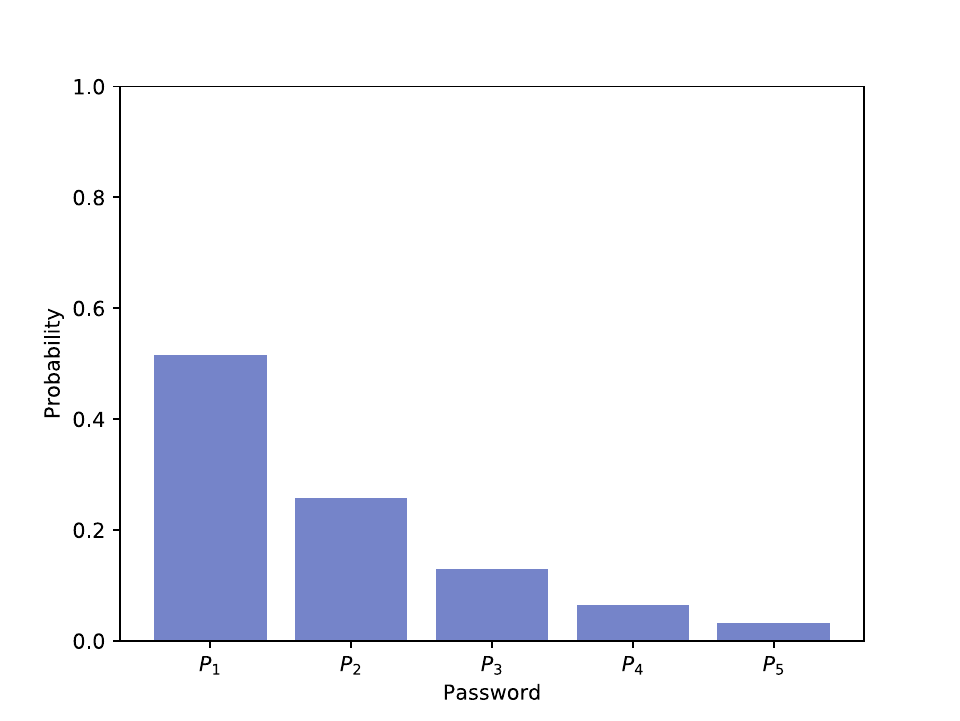}
    \caption{The simple, minimal example of a password probability distribution that we use to visualise different reselection modes in this section. Probability $D(P_{n})$ of password $P_{n}$ is $\frac{1}{2^{n}}$.}
    \label{ex:redist-eg-orig}
\end{figure}

Figure~\ref{ex:redist-eg-orig} shows a minimal example of a probability distribution derived from a hypothetical password dataset consisting of 31 user-chosen passwords, of which 5 are unique, labelled $P_1$ to $P_5$ with frequencies following the powers of 2. That is to say, the frequency $freq(P_{n})$ of password $P_{n}$ is $2^{5-n}$ and the probability $D(P_{n})$ of password $P_{n}$ is $\frac{1}{2^{n}}$.. In this section, we visualise the effect of different reselection modes on this simple example.

While it would be impossible to accurately predict this reselection process for each individual affected user, we can model certain behaviours that, if exhibited by all users, would give rise to a best, worst, or average-case security outcome. We refer to these as \textit{macrobehaviours}, and examine four of these as part of this work (though our implementation is modular, see Section~\ref{sec:skeptic}).
Given a specific \textsf{macrobehaviour}, the induced distribution obtained from imposing a password composition policy \PCP\ in a password probability distribution $D$ is denoted as:
\[
\reselection{D}{\PCP}{macrobehaviour}
\]

\subsubsection{Convergent reselection} \label{subsec:conv-resel} Every user that must reselect a password chooses the most common password that remains permitted (i.e. password choice \textit{converges} on the most common permitted password). This represents a worst-case security outcome; a larger proportion of users now have the same password, which makes the password probability distribution less uniform and the system more vulnerable to a password guessing attack containing this password.

Formally, we define this reselection mode as:
\[
\begin{array}{l}
\reselection{D}{\PCP}{convergent}(p) = \\
\hspace{2em}\begin{cases}
D(p) + \surplus{D}{\PCP} & \text{if } p=\maxPCP{\PCP}{D} \\
D(p)                     & \text{if } p\ne\maxPCP{\PCP}{D} \text{ and } \\
                         & \hspace{1em} p \in \suppPCP{\PCP}{D}\\
0                        & \text{otherwise}
\end{cases}
\end{array}
\]
Here, $\maxPCP{\PCP}{D}$ denotes the password with highest probability in $D$ that satisfies the password composition policy \PCP. This can be defined as:

\begin{align*}
\mathsf{choose}(\{\, p \,\,|\,\, & p\in\suppPCP{\PCP}{D}~~~~\wedge \\
                     & \forall p'\bullet p'\in\suppPCP{\PCP}{D} \rightarrow D(p)\ge D(p') \,\})\enspace
\end{align*}
where $\mathsf{choose}$ is non-deterministic choice of one element from the given set (which is non-empty).

\begin{figure}[ht]
    \includegraphics[width=\graphSizeMultiplier\columnwidth]{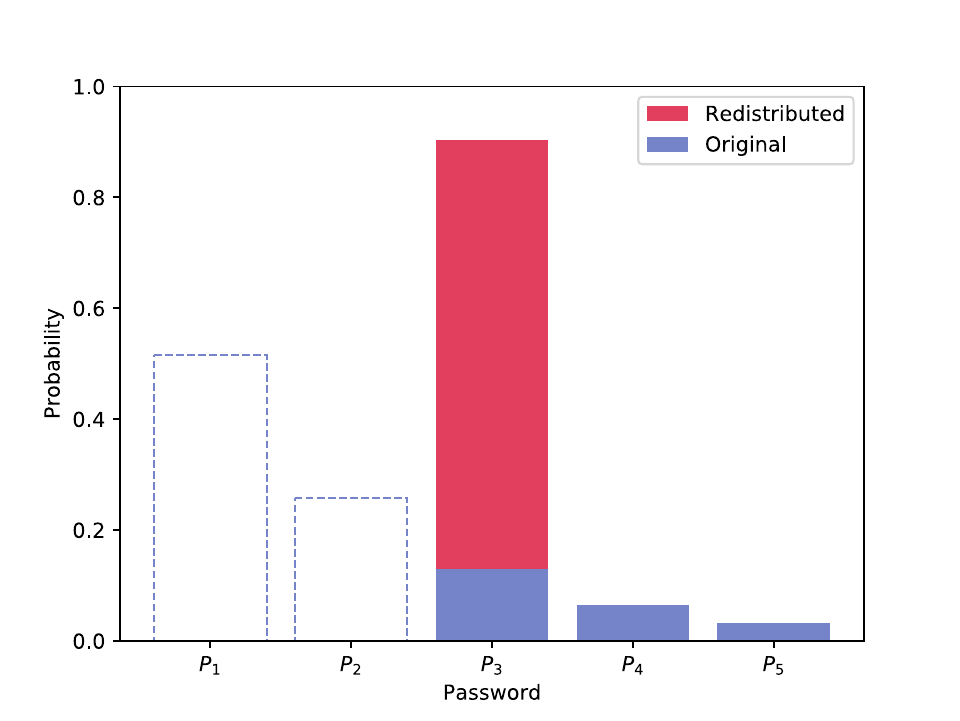}
    \caption{The redistribution of probability in convergent reselection mode under a policy prohibiting $P_1$ and $P_2$. Dotted bar outlines show the probability of prohibited passwords, and stacked bars show the redistribution of this probability.}
    \label{ex:redist-eg-conv}
\end{figure}

Figure~\ref{ex:redist-eg-conv} shows a simple example of convergent reselection applied to the example distribution shown in Figure~\ref{ex:redist-eg-orig} when a password composition policy prohibiting passwords $P_1$ and $P_2$ is applied. Note that the probability from these prohibited passwords is redistributed to the most common password $P_3$ in the dataset that remains permitted.

\subsubsection{Proportional reselection} \label{subsec:prop-resel} Every user that must reselect a password chooses a password from those remaining in a way proportional to their probabilities. This represents an average-case security outcome, with the most common remaining permitted passwords receiving the largest share of ``displaced'' users.

Formally, we define this reselection mode as:
\[
\begin{array}{l}
\reselection{D}{\PCP}{proportional}(p) = \\
\hspace{2em}\begin{cases}
\dfrac{D(p)}{1 - \surplus{D}{\PCP}} & \text{if }p \in \suppPCP{\PCP}{D} \\
0                        & \text{otherwise}
\end{cases}
\end{array}
\]

\begin{figure}[ht]
    \includegraphics[width=\graphSizeMultiplier\columnwidth]{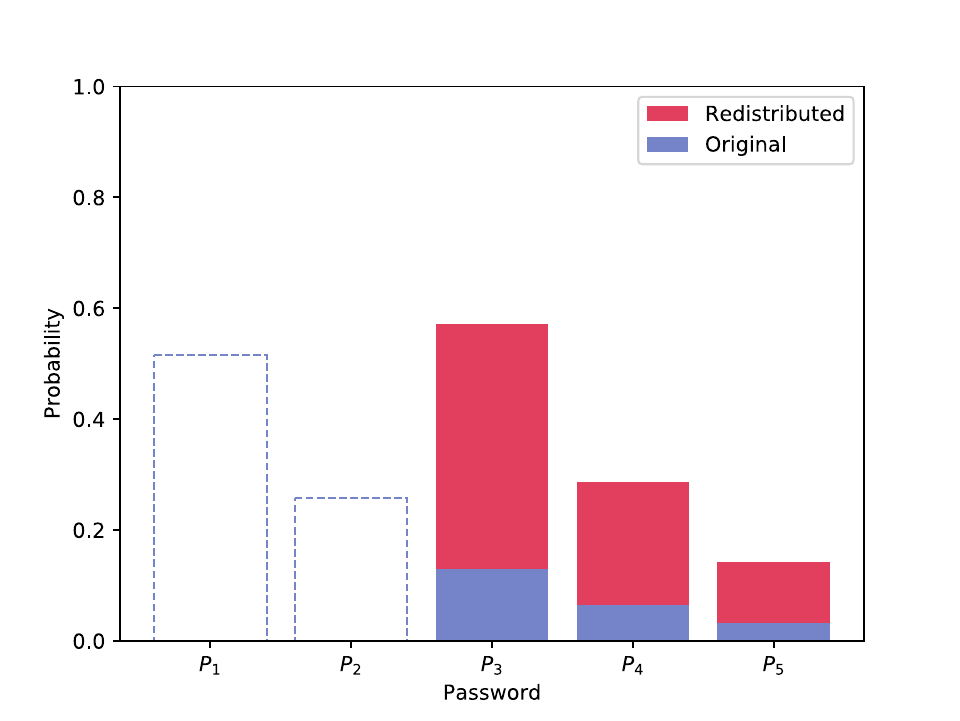}
    \caption{The redistribution of probability in proportional reselection mode under a policy prohibiting $P_1$ and $P_2$.}
    \label{ex:redist-eg-prop}
\end{figure}

Figure~\ref{ex:redist-eg-prop} shows a simple example of proportional reselection applied to the example distribution under a policy prohibiting $P_1$. and $P_2$ Note that the probability from these prohibited passwords is redistributed amongst remaining permitted passwords proportionally to their probability. 

\subsubsection{Extraneous reselection} \label{subsec:extra-resel} Every user that must reselect a password chooses a new, unique password outside the set of remaining passwords, as if they had suddenly switched to using a password manager. This represents a best-case security outcome, increasing password probability distribution uniformity to the greatest extent.

Formally, we define this reselection mode as:
\[
\begin{array}{l}
\reselection{D}{\PCP}{extraneous}(p) = \\
\hspace{2em}\begin{cases}
D(p)                     & \text{if } p \in \suppPCP{\PCP}{D} \\
\dfrac{1}{n}  & \text{if } p \in \fresh{S}{\PCP}{D}{n} \\
0                        & \text{otherwise}
\end{cases}
\end{array}
\]
where $n = \surplus{D}{\PCP} \times \magnitude{D}$ and $\fresh{S}{\PCP}{D}{n}$ is a set of $n$ new and unique passwords built from symbols in the alphabet $S$ that satisfy policy $\PCP$.
Formally, it is a set that satisfies: 
\[
|\fresh{S}{\PCP}{D}{n}| = n 
\]
and
\[
\fresh{S}{\PCP}{D}{n} = \{\, p \,|\, p \in \PCP \wedge p \not\in \supp{D} \wedge p \in S^*\,\}
\]

\begin{figure}[ht]
    \includegraphics[width=\graphSizeMultiplier\columnwidth]{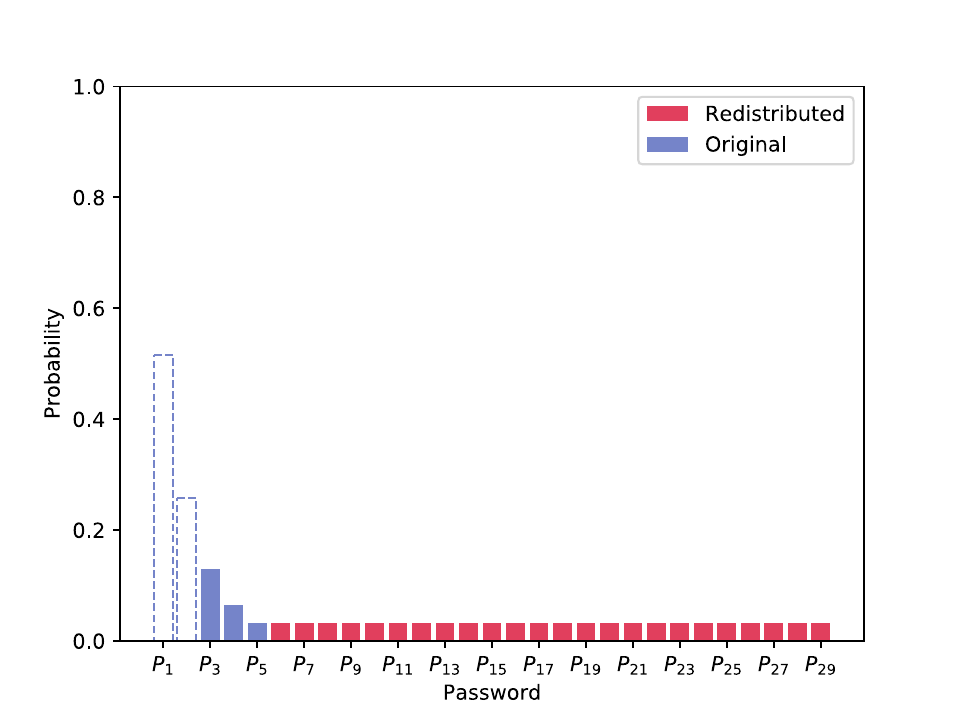}
    \caption{The redistribution of probability in extraneous reselection mode under a policy prohibiting $P_1$ and $P_2$.
    }
    \label{ex:redist-eg-extra}
\end{figure}

Figure~\ref{ex:redist-eg-extra} shows a simple example of proportional reselection applied to the example distribution under a policy prohibiting $P_1$ and $P_2$. Note that the probability from these prohibited passwords is redistributed to new, unique passwords $P_6$-$P_{29}$.

\subsubsection{Null reselection} Every user that must reselect a password simply doesn't, and never creates an account on the system. This is modelled while maintaining the probability distribution by distributing password probability completely evenly amongst all remaining permitted passwords. 

Formally, we define this reselection mode as:
\[
\begin{array}{l}
\reselection{D}{\PCP}{null}(p) = \\
\hspace{2em}\begin{cases}
D(p) + \dfrac{\surplus{D}{\PCP}}{|\suppPCP{\PCP}{D}|} & \text{if } p \in \suppPCP{\PCP}{D} \\
0                        & \text{otherwise}
\end{cases}
\end{array}
\]

\begin{figure}[ht]
    \includegraphics[width=\graphSizeMultiplier\columnwidth]{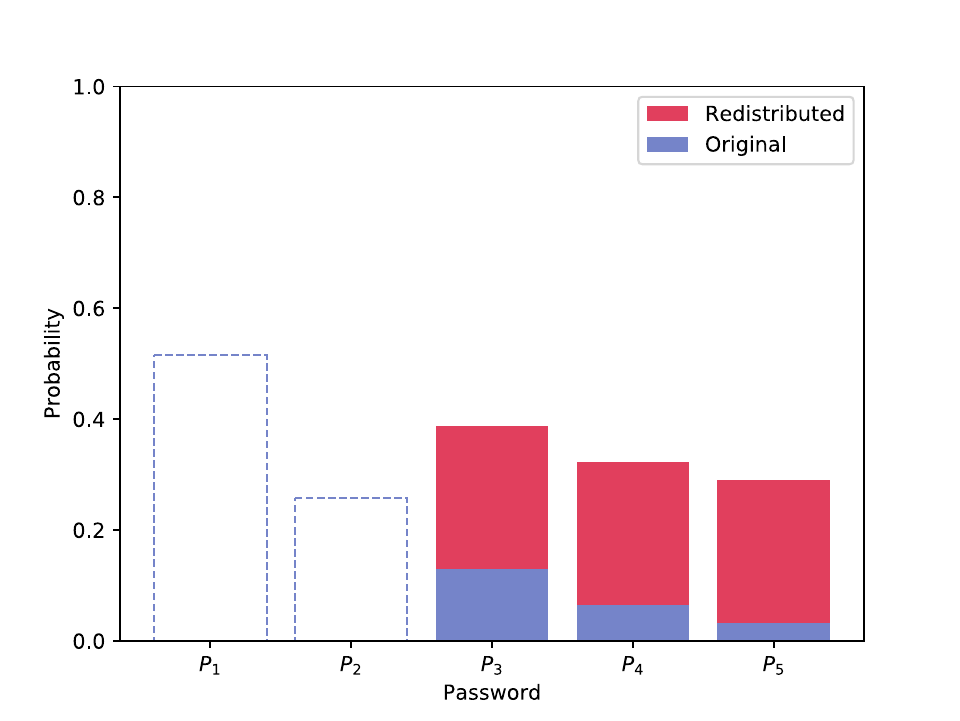}
    \caption{The redistribution of probability in null reselection mode under a policy prohibiting $P_1$ and $P_2$.
    }
    \label{ex:redist-eg-null}
\end{figure}

Figure~\ref{ex:redist-eg-null} shows a simple example of null reselection applied to the example distribution under a policy prohibiting $P_1$ and $P_2$. Note that the probability from these prohibited passwords is redistributed uniformly across remaining permitted passwords.

\subsection{Quantifying Security} \label{subsec:quantifying-security}
After transforming our probability distribution according to the policies and macrobehaviours we wish to study, we are now faced with the challenge of quantifying what it means for a distribution of user-chosen passwords to be ``secure''. To achieve this, we take advantage of the fact that more uniform distributions of user-chosen passwords are more resilient against certain password guessing attacks that rely on guessing common passwords first, due to a smaller proportion of users converging on the same popular passwords.
The notion of uniformity as a desirable property of the distribution of user-chosen passwords on a system is not new: 

\begin{itemize}
    \item Previous work by Segreti et al. \cite{segreti2017diversity} proposes password composition policies that are \textit{adaptive}---evolving over time with the express aim of increasing password diversity.
    \item Blocki et al. \cite{blocki2013optimizing} focus on maximising minimum password entropy in order to optimise password composition policies---analogous to increasing password distribution uniformity.
    \item Malone and Maher \cite{malone2012investigating} highlight that user-chosen password distributions are non-uniform, and mention that if this were not the case, attacks that rely on attempting to guess common passwords would become less effective.
\end{itemize}

We approach the problem of measuring the uniformity of password probability distributions by performing least-squares fitting of power-law equation to them of the form $y = a \times x^\alpha$. By taking $\alpha$ (the ``$\alpha$-value'' of the policy), we can compare the steepness of the fitted curves, with a shallower curve (i.e. a curve with an $\alpha$-value closer to $0$) signifying a more uniform distribution. 

\begin{figure}[ht]
    \centering
    \subfloat[Yahoo]{%
      \includegraphics[width=\graphSizeMultiplier\columnwidth]{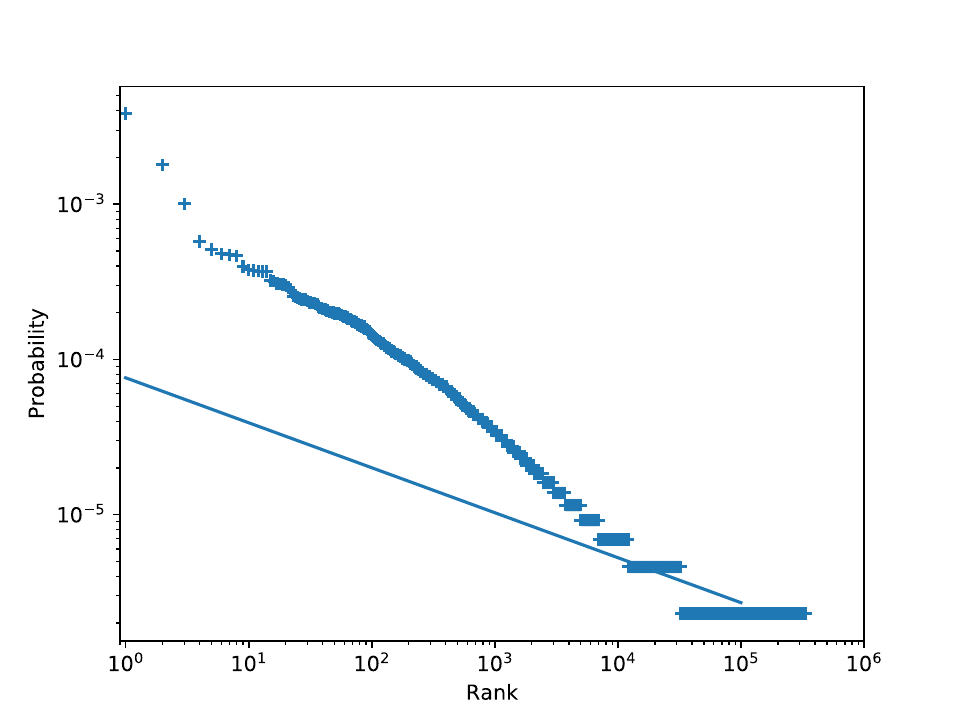}
      \label{subfig:yahoo-freqs}
    } \\
    \subfloat[Yahoo (exp. sampled)]{%
      \includegraphics[width=\graphSizeMultiplier\columnwidth]{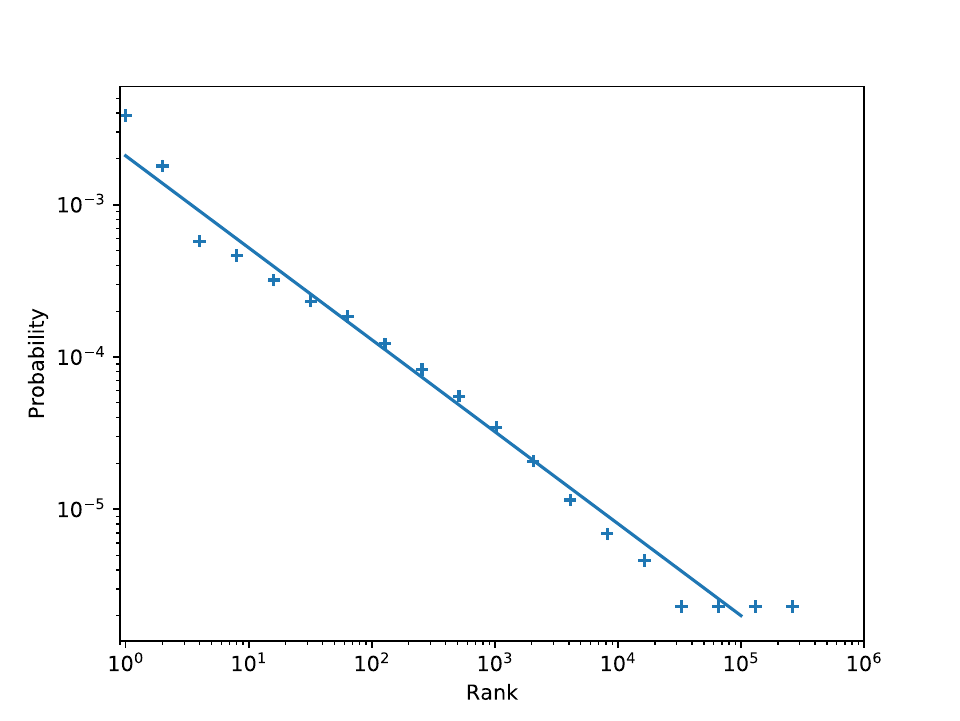}
      \label{subfig:yahoo-freqs-sampled}
    }
    \caption{The rank-probability distribution of passwords in the Yahoo dataset, with and without exponential sampling.}
    \label{fig:yahoo-freqs}
\end{figure}

This is not completely straightforward, however. Malone and Maher \cite{malone2012investigating} point out that the tendency for breached password databases to contain a high proportion of passwords with frequencies in the low-single digits causes a least-squares regression line fitted to a graph of password rank against frequency (and therefore probability) to have a slope that is too shallow (see Figure~\ref{subfig:yahoo-freqs}). Logarithmic binning of this data (that is, summing all frequencies between rank $2^n$ and $2^{n+1}$ as one data point) removes this bias, and results in a much better fit. We reproduce this result for the Yahoo data set \cite{gross2012yahoo} (which we will discuss in detail later) in Figure~\ref{subfig:yahoo-freqs-sampled}, but with an important difference---instead of summing the frequencies in each bin, we simply take every $2^{nth}$ data point and discard those in between; that is to say, we swap logarithmic binning for \textit{exponential sampling}. This similarly corrects our regression line, which now appears to interpolate the data well. Given the \textit{rank} of the probability of a password in the database between $1$ and the total number of unique passwords in the database, we can now approximate its actual probability using only the fitted equation, without requiring access to the password data itself. This allows us to justify our choice of password composition policy while avoiding the ethical concerns involved in propagating the password data that informed this choice.

\begin{figure}[ht]
    \includegraphics[width=\graphSizeMultiplier\columnwidth]{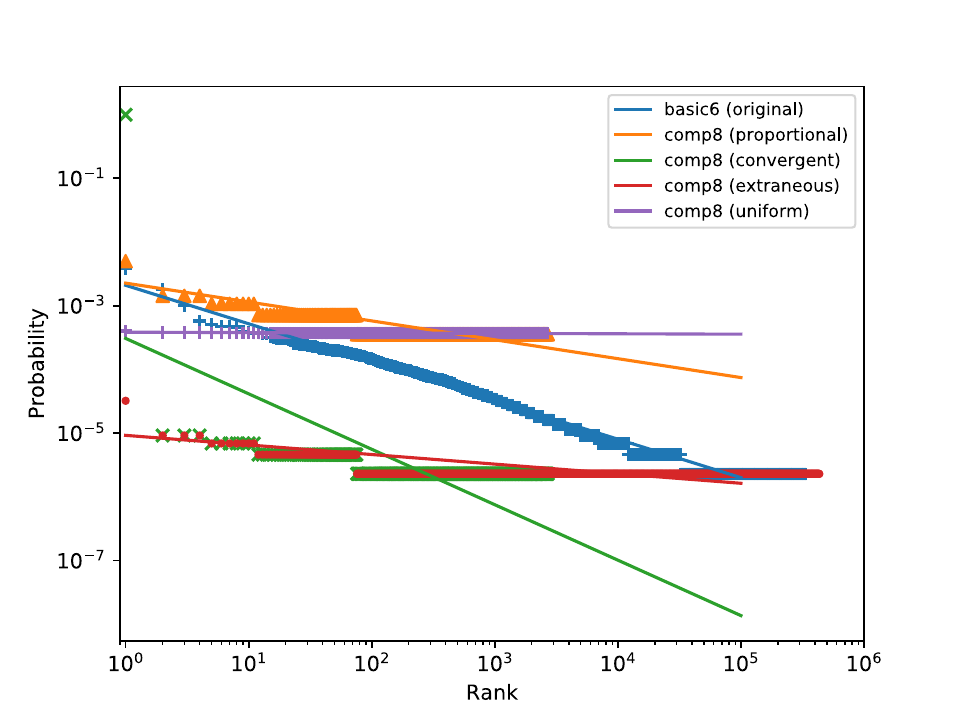}
    \caption{The original password probability distribution of the Yahoo dataset, alongside those induced by the \textit{comp8} policy under each macrobehaviour. Fitted power-law curves are also shown.}
    \label{fig:comp8-yahoo-fitting-demo}
\end{figure}

Figure~\ref{fig:comp8-yahoo-fitting-demo} shows the rank-probability distribution of the Yahoo dataset used in this study under its original policy (\textit{basic6}) and its transformations under the \textit{comp8} policy assuming each of the macrobehaviours described in Section~\ref{subsec:modelling-pwd-resel}. From the figure, it is readily apparent that different assumptions about user password reselection behaviour can lead to drastically different security outcomes for the system. While proportional, extraneous and null reselection behaviours lead to a net increase in uniformity under the \textit{comp8} policy (and therefore presumed guessing attack resistance) convergent behaviour leads to a drastic decrease.

\section{The \Skeptic{} Toolchain}\label{sec:skeptic}
We provide an implementation of the methodology in Section~\ref{sec:methodology} as a toolchain consisting of three pieces of software, designed to be used together sequentially. We name this three-part toolchain \Skeptic{}, which consists of: the metaprogramming tool \Authority{} for encoding password composition policies from within the \textit{Coq} proof assistant; the data processing tool \Pyrrho{} for redistributing password probabilities in the input set according to a password composition policy and user behaviour model; and finally \Pacpal{}, a DSL to assist system administrators in comparing and ranking password composition policies based on output from these tools. We elaborate on each of these in turn in this section.

\subsection{Policy Specification: \Authority{}} \label{subsec:authority}
Password composition policies are enforced on different systems by a diverse range of software, which may accept password policies in different encodings. It is convenient to represent these encodings as tuples containing software configuration parameters. For example, software $A$ may take a tuple $(l \in \mathbb{N}, d \in \mathbb{N})$ where $l$ is minimum password length and $d$ is the minimum number of numeric digits a password may contain; while software $B$ might take tuples $(e \in \mathbb{Q}, w \subset S^*)$ where $e$ is the minimum Shannon entropy of the password, and $w$ is a set of prohibited passwords (a ``dictionary check''). If we wish to compare one of each of these tuples, we must first obtain them in a uniform (i.e. normalised) encoding.

To achieve this, we take advantage of the fact that any password composition policy is necessarily a predicate on passwords (i.e. a function with type $Password \rightarrow \mathbb{B}$). With this in mind, we can obtain a uniform representation of password composition policies regardless of the software they were encoded for by devising a function to decode them to a Boolean normal form. For software $A$ for example, we might devise the function in Equation~\ref{eq:norm-func-a} which will transform a password composition policy encoded for this software into a predicate in conjunctive normal form.

\begin{equation} \label{eq:norm-func-a}
    norm_A(l, d) = \lambda s. length(s) \geq l \wedge digits(s) \geq d
\end{equation}

Even though software $B$ takes a different configuration tuple, we need only specify the normalisation function in Equation~\ref{eq:norm-func-b} for tuples of this type in order to obtain a password composition policy predicate in the same representation. 

\begin{equation} \label{eq:norm-func-b}
    norm_B(e, w) = \lambda s. shannon(s) \geq e \wedge s \notin w
\end{equation}

Normalisation functions specified in this way are amenable to formal verification, not only with respect to their correctness (i.e. their conversion of software-specific configuration tuples to predicates) but also desirable properties of the predicates they generate. For instance, we can show that a policy mandating a minimum password length of 16 encoded for software $A$ as configuration tuple $(16, 0)$ and normalised to policy predicate $\phi$ confers immunity to a guessing attack consisting of passwords in an arbitrary set of guesses $G$ by showing the universal quantification in Equation~\ref{eq:simple-immunity-eg} holds.

\begin{equation} \label{eq:simple-immunity-eg}
    \phi = norm_A(16, 0) \quad\quad\quad \forall g \in G. \neg\phi(g)
\end{equation}

\Authority{} is a metaprogramming utility\footnote{We make \Authority{} available as open-source software: \\ 
\url{https://github.com/sr-lab/skeptic-authority-template/}}
that enables the interactive modelling of password composition policies for arbitrary software, generating a \textit{Coq} project. From the \textit{Coq} interactive theorem proving environment, it is then possible to both specify and verify the correctness of a normalisation function for transforming password composition policies encoded as software-specific tuples into predicates (see Section~\ref{subsec:specifying-pcps}) as well as desirable properties of the password composition policies themselves, such as immunity to certain guessing attacks that malware uses to propagate (see Section~\ref{subsec:pol-immunity}). This command-line utility asks the user a series of questions, guiding them through this process:

\begin{enumerate}
    \item They are first asked to specify the name, type and description of each member of the type of software-specific configuration tuple they wish to model.
    \item Then, they may optionally specify an arbitrary number of different password composition policies encoded as tuples of this type by specifying policy names and tuple values.
    \item A ready-to-use \textit{Coq} project is then generated according to the user's specifications. All that remains is for the user to manually specify the normalisation function (see Section~\ref{subsec:specifying-pcps}) to convert the password composition policy tuples into predicates.
\end{enumerate}

\begin{figure}[ht]
    \centering
    \includegraphics[width=\figureSizeMultiplier\columnwidth]{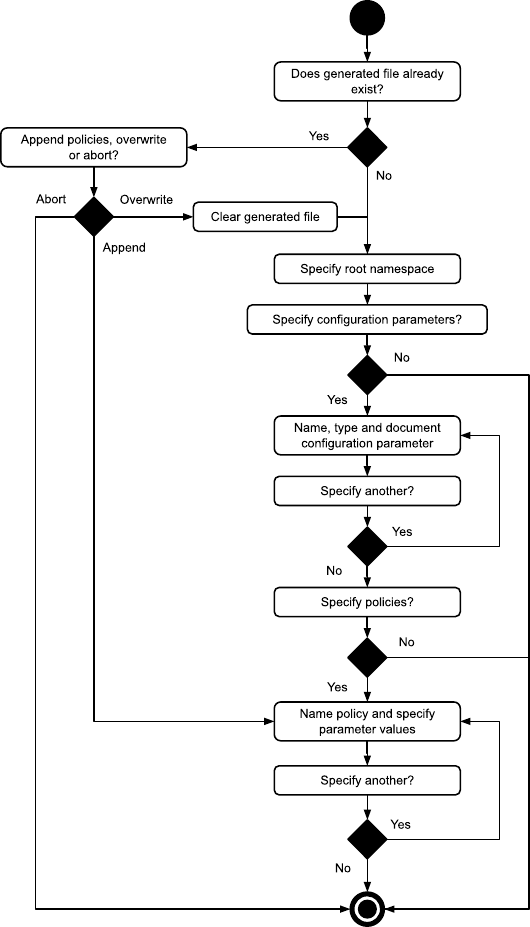}
    \caption{A simplified overview of the logical flow of a run of the \Authority{} utility.}
    \label{fig:authority_flow}
\end{figure}

For a more detailed overview of the operation of \Authority{}, see the flow diagram in Figure~\ref{fig:authority_flow}. Included in the generated \textit{Coq} project are various tools designed to streamline the process of proving desirable properties about the password composition policies encoded using the tool, including a trie implementation for high-performance dictionary checks, a pre-built notion of immunity and a simple \texttt{simulate} tactic that can be used to prove properties about password composition policies with respect to smaller guessing attacks by simple simulation.

A central feature of \Authority{} is that is can be used by \Pyrrho{}, the next utility in the \Skeptic{} toolchain, to filter large sets of real-world user password data in order to model changes in the distribution of passwords under different password composition policies and user macrobehaviours. Password composition policies can therefore be modelled from within \textit{Coq}, and used directly for this filtration step. \Authority{} achieves this by making use of the \textit{Coq.io} \cite{claret2015coq} library for writing IO-enabled programs in \textit{Coq}, and communicating with \Pyrrho{} (which is written in Python for optimal performance) via its standard output stream.

\subsection{Password Reselection: \Pyrrho{}} \label{subsec:pyrrho}
\begin{figure}[ht]
    \centering
    \includegraphics[width=\columnwidth]{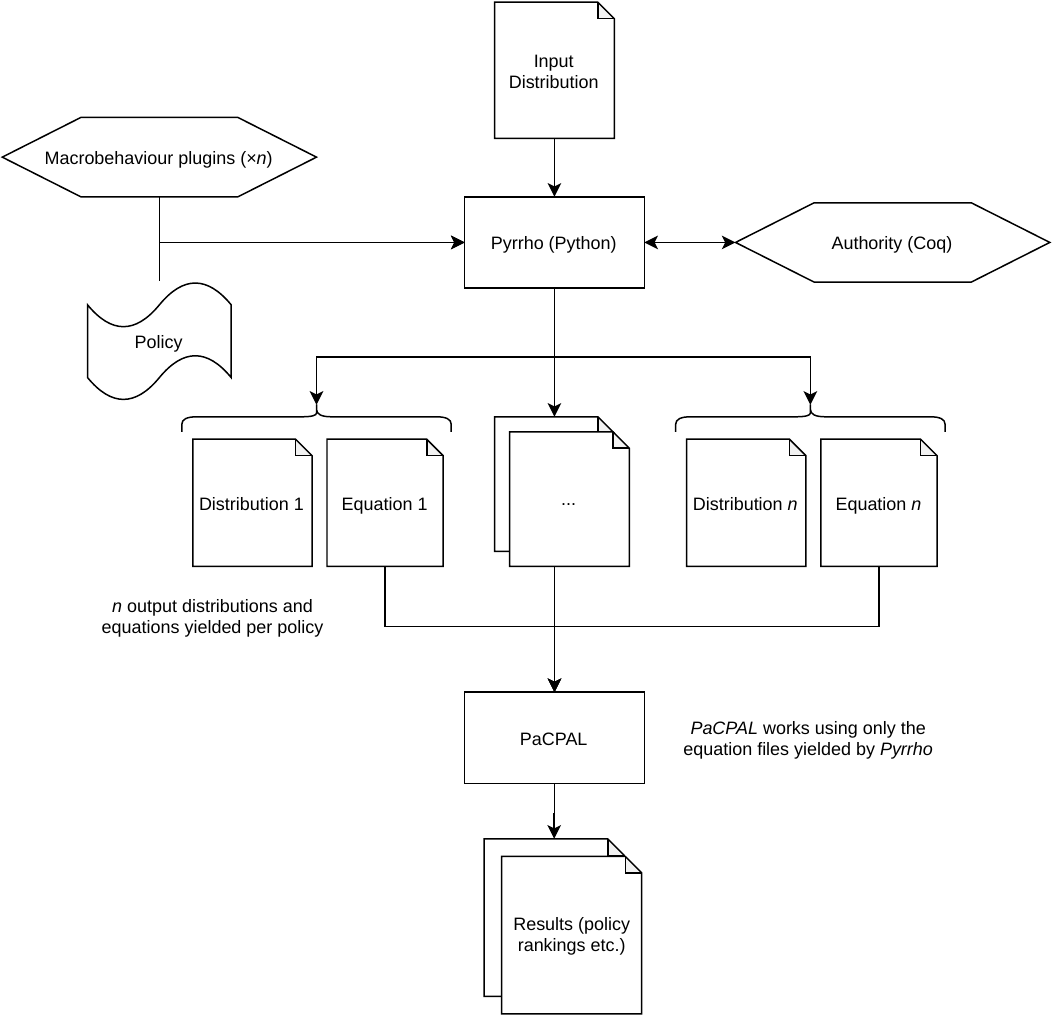}
    \caption{An overview of the function of \Skeptic{}. Arrows indicate the direction of data flow.}
    \label{fig:skeptic-overview}
\end{figure}

\Pyrrho{} lies at the core of the \Skeptic{} toolchain, a software tool\footnote{We make \Pyrrho{} available as open-source software: \\ 
\url{https://github.com/sr-lab/pyrrho}}
written in Python that handles the transformation of password probability distributions derived from real-world datasets according to password composition policies and assumptions about user behaviour (i.e. the macrobehaviours discussed in Section~\ref{subsec:modelling-pwd-resel}). Figure~\ref{fig:skeptic-overview} shows an overview of the \Skeptic{} toolchain, and the position of \Pyrrho{} within it, with arrows indicating the direction of data flow between tools.

The utility is parametric on a password probability distribution derived from a real-world leaked password dataset. Password probabilities are then redistributed according to a password composition policy (interpreted by \Authority{}), producing output distributions under each supported macrobehaviour. Its architecture is modular, allowing user-specified macrobehaviours to be plugged in without any modification to the core of the tool. The \Pyrrho{} plugin corresponding to the proportional password reselection macrobehaviour from  Section~\ref{subsec:modelling-pwd-resel} is shown in Figure~\ref{fig:python-plugin}. Here, \texttt{total} is the sum of all probabilities in the distribution before filtration (which should be $\approx1$), \texttt{surplus} is the sum of the probabilities of all filtered passwords, and \texttt{df} is the data frame representing the password probability distribution to process.

\begin{figure}[ht] \codeSampleFontSize
    \centering
    \begin{minted}{python}
def reselect (total, surplus, df):
    divisor = total - surplus
    df['probability'] /= divisor
    return df
    \end{minted}
    \caption{The proportional password reselection macrobehaviour from Section~\ref{subsec:modelling-pwd-resel} encoded in Python as a plugin for \Pyrrho{}.}
    \label{fig:python-plugin}
\end{figure}

\Pyrrho{} additionally performs power-law curve fitting to the altered password probability distributions in order to quantify their uniformity (see Section~\ref{subsec:quantifying-security}), storing the resulting equations encoded as JSON files alongside them. It is these JSON files that can be used to compare and rank policies from the \Pacpal{} DSL (see Section~\ref{subsec:pacpal}).

While \Pyrrho{} is primarily designed to be used alongside password composition policies encoded in \textit{Coq} using \Authority{}, the inter-process communication involved between the two utilities makes processing large datasets a time-consuming process. For applications where the ability to reason about password composition policies from within \textit{Coq} is less important, \Pyrrho{} also supports \textit{Pure Python Mode}, in which all dataset filtration with respect to a password composition policy is kept within \Pyrrho{} itself. The result is a utility which runs on the order of $2.75$ times faster (see Section~\ref{subsec:experimental-setup}), but at the expense of the flexibility of password composition policy encoding and reasoning that comes with using \Authority{}, as \textit{Pure Python Mode} supports only a limited set of password composition policy rules.

\subsection{Result Extraction: \Pacpal{}} \label{subsec:pacpal}
While the data produced by \Pyrrho{} is ostensibly all we need to be able to assess the relative security of password composition policies under our assumptions, the nuance of this data is of comparatively little interest to professionals working in  an applied setting (system administrators, for example). 

\begin{figure}[ht] \codeSampleFontSize
    \begin{minted}{text}
# Load three equations produced by Pyrrho.
load linkedin-basic16-proportional.json as li_b16
load linkedin-2word16-proportional.json as li_2w16
load linkedin-3class12-proportional.json as li_3c12

# Assert that one policy is better than another.
assert li_2w16 better li_b16 

# Build group to rank.
group linkedin_ranking
add li_b16 to linkedin_ranking as basic16
add li_2w16 to linkedin_ranking as 2word16
add li_3c12 to linkedin_ranking as 3class12

# Print group in ranked order (worst to best):
rank linkedin_ranking
\end{minted}
    \caption{A piece of example \Pacpal{} code, demonstrating ranking of policies based on fitted power-law equations.}\label{fig:pacpal-skeptic-eg}
\end{figure}

Users such as this are likely to be far more interested in choosing the most secure password composition policy for their use-case than in the data itself. \Pacpal{}\footnote{We make \Pacpal{} available as open-source software: 
\url{https://github.com/sr-lab/skeptic-lang}}
is an assertion language permitting power-law equations generated by \Pyrrho{} to be loaded, named, grouped, compared and ranked, and is designed to assist end-users in putting \Skeptic{} to work practically in their organisations, leveraging the well-documented usability benefit seen with domain-specific languages when compared to their general-purpose counterparts \cite{bariic2012usability}. An example piece of \Pacpal{} code is shown in Figure~\ref{fig:pacpal-skeptic-eg} in which three fitted power-law equation files produced by \Pyrrho{} are loaded, bound to names, added to a group and ranked. The ranking will then be displayed to the user. Also present is a \texttt{better} assertion which will display an error to the user in the case that this relationship does not hold. We employ \Pacpal{} to produce the rankings of all \numPolices~password policies used in this study in Section~\ref{subsec:ideal-policies}.

\section{Evaluation}\label{sec:evaluation}
In this section, we demonstrate the validity of our approach by replicating results from previous literature across different evaluation methodologies. Specifically, we use the \Skeptic{} toolkit to replicate results from the study by Shay et al. \cite{shay2016designing} that uses real participants recruited via Amazon Mechanical Turk (see Section~\ref{subsec:replic-shay}), and the study by Weir et al. \cite{weir2010testing} that draws on large leaked password datasets (see Section~\ref{subsec:replic-weir}). In Section~\ref{subsec:pol-immunity}, we demonstrate the advantages of the \Authority{} \textit{Coq} metaprogramming utility (see Section~\ref{subsec:authority}) by proving that certain policies confer immunity to password guessing attacks by some common botnet worms from within the proof assistant itself.

\subsection{Experimental Setup} \label{subsec:experimental-setup}
The password probability distribution processing (via \Pyrrho{}) for this experiment was conducted on a cluster of 14 cloud-based virtual machines, each with 6 Intel\textregistered{} Xeon\textregistered{} CPUs at 1.80GHz, 16GB of RAM and 320GB of hard disk space running 64-bit Ubuntu 18.04.3 (LTS). Times taken by \Pyrrho{} to process each dataset studied in this work under each policy and macrobehaviour we investigate are shown in Table~\ref{tbl:pyrrho-proc-times}.

\begin{table}[ht] \tableFontSize
    \caption{Time taken for \Pyrrho{} to process probability distributions for each of the datasets, policies and macrobehaviours investigated.}
    \label{tbl:pyrrho-proc-times}
    \centering
    \begin{threeparttable}
        \begin{tabular}{|l|l|l|l|}
            \hline
            Dataset   & Time (s)  & Uniq. passwords & Time/password \\ \hline
            Yahoo     & 17,817    & 337,168         & 0.0528        \\ \hline 
            Yahoo*    & 6,466     & 337,168         & 0.0192        \\ \hline 
            RockYou*  & 339,708   & 14,308,965      & 0.0237        \\ \hline 
            LinkedIn* & 1,741,996 & 60,489,959      & 0.0288        \\ \hline 
        \end{tabular}
        \vspace{0.5em} 
        \begin{tablenotes}
            \footnotesize
            \item * Computed in \textit{\Pyrrho{}'s} pure Python mode for reasons of performance.
        \end{tablenotes}
    \end{threeparttable}
\end{table}

\subsection{Replication of Results: Shay et al.}\label{subsec:replic-shay}
Shay et al.~\cite{shay2016designing} ranked the effectiveness of 8 different password composition policies under a password guessing attack at two different magnitudes---$10^{6}$ guesses and $10^{14}$ guesses. These two thresholds are suggested by Flor\^encio et al.~\cite{florencio2014password} as being representative of the cutoff points of contemporary online (i.e. against a live service) and offline (i.e. against a compromised password hash) guessing attacks respectively. Passwords were chosen by humans under each policy using Amazon Mechanical Turk and the attack was multimodal using both a trained, targeted probabilistic context-free grammar (PCFG) \cite{weir2009password,kelley2012guess} and the \textit{Password Guessability Service} (PGS) \cite{ur2015measuring}. Table~\ref{tbl:shay-results} contains an overview of these results.

\begin{table}[ht] \tableFontSize
\caption{The results obtained by Shay et al.~\cite{shay2016designing} for passwords collected under 8 different password composition policies at both attack magnitudes.}
\begin{tabular}{|l|l|l|l|l|}
\hline
         & \multicolumn{2}{c|}{$10^{6}$ guesses} & \multicolumn{2}{c|}{$10^{14}$ guesses} \\ \hline
Policy   & Cracked (\%)          & Rank          & Cracked (\%)           & Rank          \\ \hline
comp8    & 2.2                   & 3             & 50.1                   & 7             \\ \hline
basic12  & 9.1                   & 8             & 52                     & 8             \\ \hline
basic16  & 7.9                   & 7             & 29.7                   & 4             \\ \hline
basic20  & 5.6                   & 6             & 16.4                   & 2             \\ \hline
2word12  & 3.4                   & 5             & 46.6                   & 6             \\ \hline
2word16  & 1.1                   & 1             & 22.9                   & 3             \\ \hline
3class12 & 3.2                   & 4             & 36.8                   & 5             \\ \hline
3class16 & 1.2                   & 2             & 13.8                   & 1             \\ \hline
\end{tabular}
\label{tbl:shay-results}
\end{table}

We attempted to replicate these results using the \Skeptic{} toolkit. For each of our 3 datasets, and each of the 4 studied macrobehaviours, we redistributed probability according to each policy in Table~\ref{tbl:shay-results}. We then obtained the $\alpha$ values yielded by fitting power-law curves to the resulting distributions using the methodology described in Section~\ref{subsec:quantifying-security}. In order to quantify how closely our results reflect the rankings from Shay et al.~\cite{shay2016designing} we plotted the percentage of passwords cracked under each policy in Shay et al.~\cite{shay2016designing} against the $\alpha$-values we obtained using our methodology and calculated the Pearson correlation coefficient $\rho$. A value closer to $-1$ indicates that more uniform distributions (i.e. a less negative $\alpha$-value) are more strongly correlated with a lower percentage of cracked passwords according to Shay et al.~\cite{shay2016designing}, while a value closer to $1$ indicates the opposite. A value of $0$ indicates no correlation. The complete set of correlation coefficients and their mean values across datasets $\bar{\rho}$ can be found in Table~\ref{tbl:shay-corr-co-14}, while an example visualisation using the LinkedIn dataset only is shown in Figure~\ref{fig:linkedin-offline-rho-plot-perc-proportional}. Complete results are shown in the Appendix (Table~\ref{tbl:complete-results}). 

\begin{table}[ht] \tableFontSize
    \caption{Pearson correlation coefficients of percentage of passwords cracked under different polices by Shay et al.~\cite{shay2016designing} at $10^{14}$ guesses against $\alpha$-values yielded by \Skeptic{}.}
    \begin{threeparttable}
        \begin{tabular}{|l|l|l|l|l|}
            \hline
            Mode         & Yahoo  & RockYou & LinkedIn* & $\bar{\rho}$ \\ \hline
            Proportional & -0.661 & -0.591  & -0.929    & -0.727  \\ \hline
            Convergent   & 0.882  & -0.069  & 0.615     & 0.476   \\ \hline
            Extraneous   & -0.722 & -0.689  & -0.952    & -0.788  \\ \hline
            Null         & -0.550 & -0.565  & -0.884    & -0.666   \\ \hline
        \end{tabular}
        \vspace{0.5em} 
        \begin{tablenotes}
            \footnotesize
            \item * Visualised in Figure~\ref{fig:linkedin-offline-rho-plot-perc-proportional}.
        \end{tablenotes}
    \end{threeparttable}
\label{tbl:shay-corr-co-14}
\end{table}

\begin{figure}[ht]
    \centering
    \includegraphics[width=0.9\columnwidth]{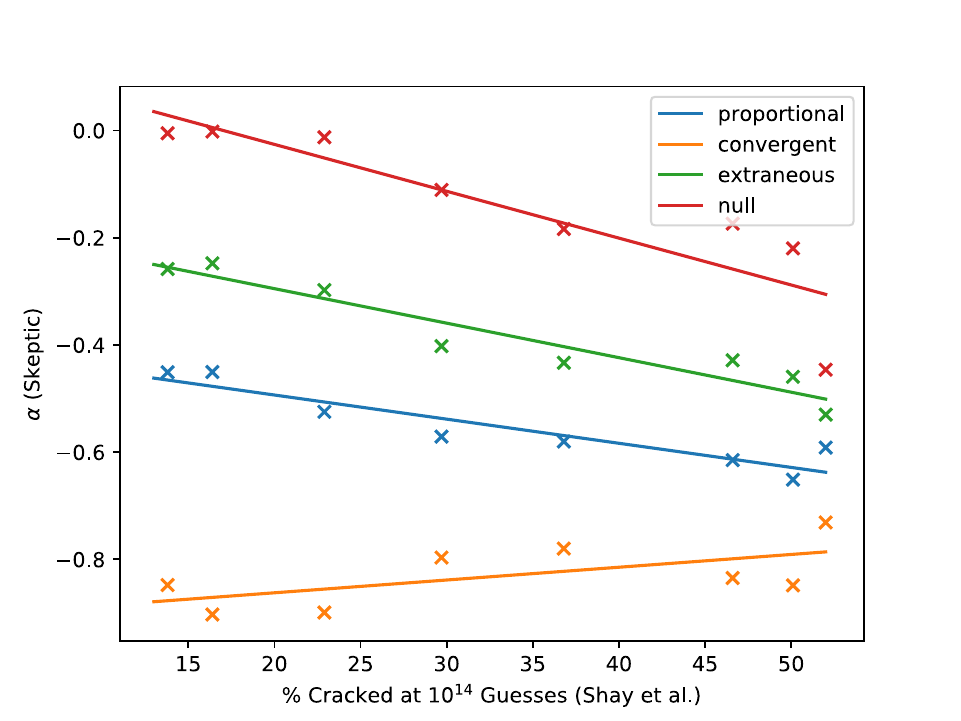}
    \caption{Percentage of passwords cracked by Shay et al.~\cite{shay2016designing} at $10^{14}$ guesses against $\alpha$-values yielded by \Skeptic{} for the LinkedIn dataset in each reselection mode.}
    \label{fig:linkedin-offline-rho-plot-perc-proportional}
\end{figure}

From Table~\ref{tbl:shay-corr-co-14}, it is apparent that $\alpha$-values for proportional, extraneous and null macrobehaviours tend to correlate well with the empirical results from Shay et al.~\cite{shay2016designing}. Using thresholds proposed by Evans~\cite{evans1996straightforward}, correlation strengths range from \textit{moderate} ($0.40 \leq |\rho| \leq 0.59$) to \textit{very strong} ($0.80 \leq |\rho| \leq 1.0$) for each of these macrobehaviours across all 3 datasets, with an average correlation strength of \textit{strong} ($0.60 \leq |\rho| \leq 0.79$). By contrast, the convergent macrobehaviour tends to show a correlation in the opposite direction, with less uniform distributions being associated with lower percentages of cracked passwords. This suggests the convergent macrobehaviour is a poor model of how users actually reselect passwords in response to password composition policies. 

\begin{figure}[ht]
    \centering
    \includegraphics[width=0.9\columnwidth]{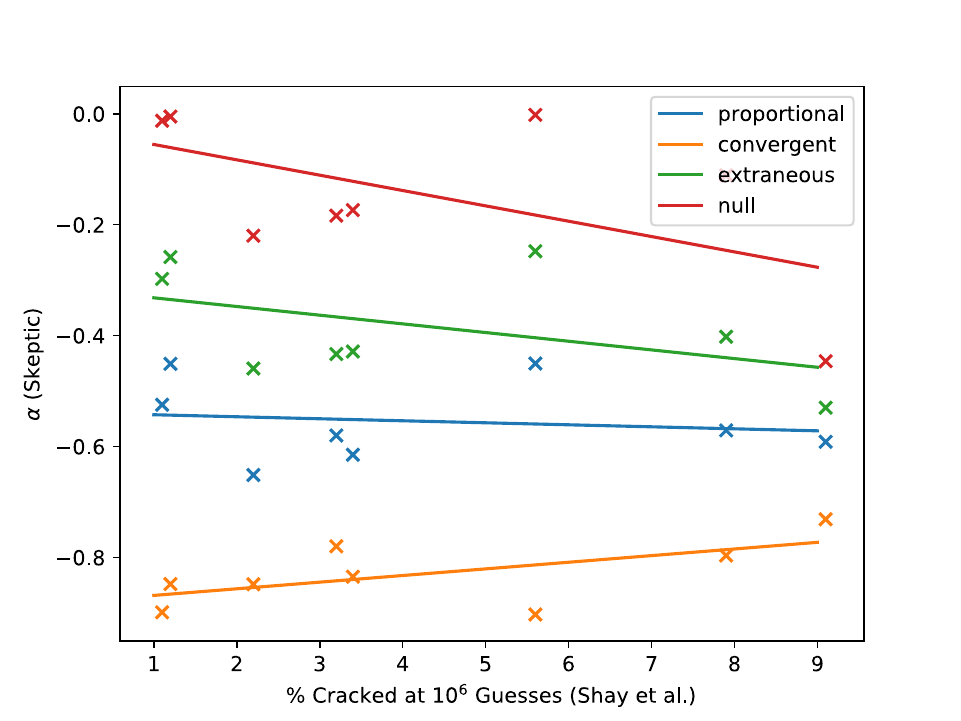}
    \caption{Percentage of passwords cracked by Shay et al.~\cite{shay2016designing} at $10^{6}$ guesses against $\alpha$-values yielded by \Skeptic{} for the LinkedIn dataset in each reselection mode.}
    \label{fig:linkedin-online-rho-plot-perc-proportional}
\end{figure}

We found $\alpha$-values yielded by \Skeptic{} to correlate slightly less closely with the percentage of passwords cracked by the smaller online-range guessing attack from Shay et al.~\cite{shay2016designing} (see Table~\ref{tbl:shay-corr-co-6} and Figure~\ref{fig:linkedin-online-rho-plot-perc-proportional}). We imagine that this is due to the success of smaller guessing attacks being more dependent on the dataset they are performed against. It is also possible that the multimodal attack employed by Shay et al.~\cite{shay2016designing} is causing guessing attacks at lower magnitudes to be more effective against passwords created under different password composition policies than at higher magnitudes.

\begin{table}[ht] \tableFontSize
    \caption{Pearson correlation coefficients of percentage of passwords cracked under different polices by Shay et al.~\cite{shay2016designing} at $10^{6}$ guesses against $\alpha$-values yielded by \Skeptic{}.}
    \begin{threeparttable}
        \begin{tabular}{|l|l|l|l|l|}
            \hline
            Mode         & Yahoo  & RockYou & LinkedIn* & $\bar{\rho}$ \\ \hline
            Proportional & -0.866 & -0.676  & -0.149    & -0.564 \\ \hline
            Convergent   & 0.217  & -0.181  & 0.615     & 0.217 \\ \hline
            Extraneous   & -0.830 & -0.808  & -0.462    & -0.700 \\ \hline
            Null         & -0.684 & -0.797  & -0.558    & -0.680 \\ \hline
        \end{tabular}
        \vspace{0.5em} 
        \begin{tablenotes}
            \footnotesize
            \item * Visualised in Figure~\ref{fig:linkedin-online-rho-plot-perc-proportional}.
        \end{tablenotes}
    \end{threeparttable}
\label{tbl:shay-corr-co-6}
\end{table}

The observation that the proportional, null and extraneous macrobehaviours offer a more accurate picture of user password reselection than convergent reselection is encouraging, because each of these represents a net increase (rather than decrease) in the uniformity of the password distribution on the system. This leads us to the conclusion that implementation of stricter password composition policies does, in general, lead to an increase in the resistance of a system to password guessing attacks. Noteworthy, however, are the outlying $\rho$ values for the convergent macrobehaviour on the RockYou dataset (see Tables~\ref{tbl:shay-corr-co-14} and \ref{tbl:shay-corr-co-6}), which seem to indicate that user password reselection behaviour for this dataset more closely resembles the convergent macrobehaviour. This is possibly due to the age of this dataset in comparison to the others (2009 vs. 2012) and consequently less secure password reselection behaviours by users of that system. This may be demographics and use-case-related, with RockYou being an online gaming service that may have had a higher proportion of younger users less adept at picking secure passwords, or users who place comparatively little value on online gaming accounts compared to those tied directly to their professional or social lives (e.g. the LinkedIn professional social networking site or Yahoo Voices online publishing platform).

\paragraph{\textbf{Findings}} Overall, \Skeptic{} produces $\alpha$-values, and therefore password composition policy rankings, that are strongly correlated with the results obtained by Shay et al. \cite{shay2016designing} from real human users recruited to create passwords under various password composition policies. This is particularly true when attack magnitude is greater (e.g. offline attacks) as opposed to smaller attacks in the online range which are more sensitive to the specific password distribution they are conducted against. Because \Skeptic{} takes password distribution uniformity as a measure of security, and thus is attack-independent, this is to be expected. This uniformity-based methodology employed by \Skeptic{} is an accurate measure of general resistance to password guessing attacks, but a considerably poorer measure of resistance to specific, targeted attacks tailored with a specific password distribution in mind.

\subsection{Replication of Results: Weir et al.}\label{subsec:replic-weir}
We next turn our attention to a study by Weir et al. \cite{weir2010testing} which draws on leaked password datasets in order to attempt to determine password composition policy effectiveness, rather than collecting passwords from humans themselves under those policies.

\begin{table}[ht] \tableFontSize
\caption{An approximation of the results obtained by Weir et al.~\cite{weir2010testing} for passwords obtained under 12 different password composition policies by filtering their target dataset.}
\begin{tabular}{|l|l|l|}
\hline
         & \multicolumn{2}{l|}{$5 \times 10^{4}$ guesses} \\ \hline
Policy   & Cracked (\%) & Rank           \\ \hline
basic7   & 26.06        & 12             \\ \hline 
basic8   & 23.16        & 11             \\ \hline 
basic9   & 18.98        & 10             \\ \hline 
basic10  & 13.85        & 8              \\ \hline 
upper7   & 13.89        & 9              \\ \hline 
upper8   & 10.71        & 7              \\ \hline 
upper9   & 7.71         & 6              \\ \hline 
upper10  & 5.72         & 4              \\ \hline 
symbol7  & 6.92         & 5              \\ \hline 
symbol8  & 5.57         & 3              \\ \hline 
symbol9  & 4.76         & 2              \\ \hline 
symbol10 & 3.28         & 1              \\ \hline 
\end{tabular}
\label{tbl:weir-pol-ranks}
\end{table}

This work, among other results, presents the percentage of passwords cracked at $50,000$ guesses under 4 different password length thresholds (7, 8, 9 and 10) and 3 different character requirements (none, at least one uppercase and at least one symbol). Both the target passwords and the attack were drawn from separate subsets of the same RockYou dataset \cite{cubrilovich2009rockyou} we make use of in this work. We present an approximation of results from \cite{weir2010testing} in Table~\ref{tbl:weir-pol-ranks}, obtained using a plot digitiser\footnote{We used \textit{WebPlotDigitizer}: \url{https://github.com/ankitrohatgi/WebPlotDigitizer}} from the visualisations in the work.

\begin{table}[ht] \tableFontSize
    \caption{Pearson correlation coefficients of password policy ranks from \cite{weir2010testing} at $5 \times 10^{4}$ guesses against $\alpha$-values yielded by \Skeptic{}.}
    \begin{threeparttable}
        \begin{tabular}{|l|l|l|l|l|}
        \hline
        Mode         & Yahoo  & RockYou & LinkedIn  & Mean   \\ \hline
        Proportional & -0.884 & -0.916  & -0.885    & -0.895 \\ \hline
        Convergent   & 0.686  & -0.657  & 0.234     & 0.089  \\ \hline
        Extraneous   & -0.955 & -0.951  & -0.969    & -0.958 \\ \hline
        Null         & -0.953 & -0.945  & -0.967    & -0.955 \\ \hline
        \end{tabular}
    \end{threeparttable}
\label{tbl:weir-correl-co}
\end{table}

\begin{figure}[ht]
    \centering
    \includegraphics[width=0.9\columnwidth]{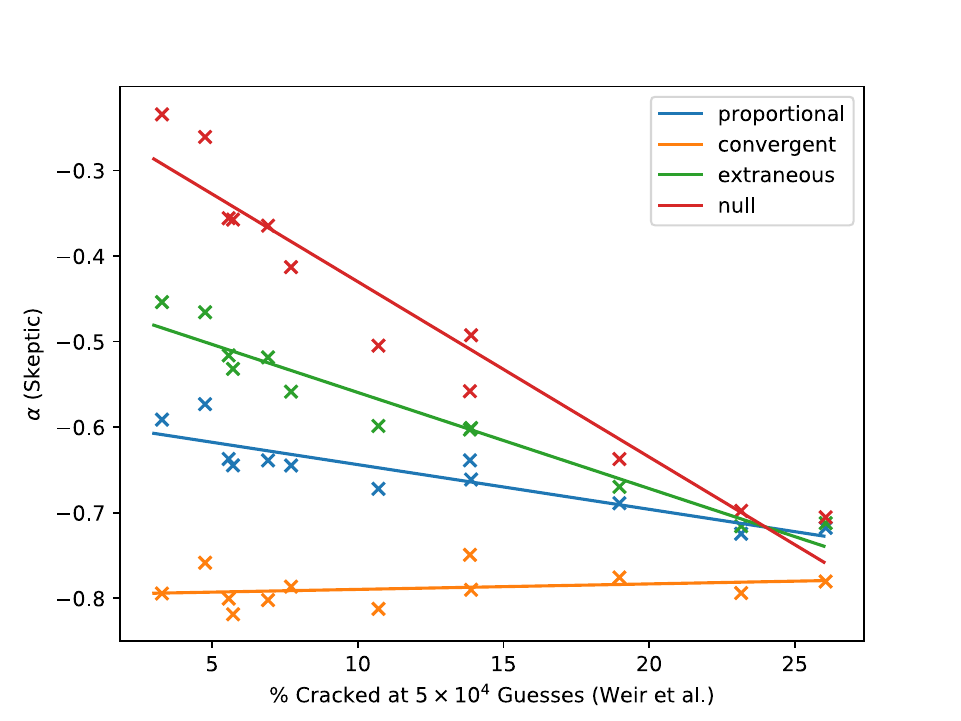}
    \caption{Percentage of passwords cracked in Weir et al.~\cite{weir2010testing} at $5 \times 10^{4}$ guesses against $\alpha$-values yielded by \Skeptic{} for the LinkedIn dataset in each reselection mode.}
    \label{fig:linkedin-weir-rho-plot-perc-proportional}
\end{figure}

Under these policies, \Skeptic{} produces $\alpha$-values that correlate very strongly with the percentage of passwords guessed by Weir et al.~\cite{weir2010testing} in proportional, extraneous, and null reselection modes (see Table~\ref{tbl:weir-correl-co}).
The $\alpha$-values for the LinkedIn dataset under each policy and macrobehaviour studied are plotted against percentages of passwords cracked by Weir et. al~\cite{weir2010testing} in Figure~\ref{fig:linkedin-weir-rho-plot-perc-proportional}.

In convergent reselection mode, \Skeptic{} is much less accurate for the Yahoo and LinkedIn datasets, but retains a strong correlation for the RockYou set. We speculate that this is for the same dataset-specific reasons as presented in Section~\ref{subsec:replic-shay} but more pronounced due to the use of the same dataset in both that work, and this one.

\paragraph{\textbf{Findings}} \Skeptic{} produces $\alpha$-values and policy rankings that are very strongly correlated with results obtained by Weir et al.~\cite{weir2010testing} from large sets of revealed password data.

\subsection{Policy Ranking}\label{subsec:ideal-policies}
If we wish to make an informed choice of password composition policy, one way we might go about this is to rank our candidates in order from most to least secure and use the resulting ranking to make our decision. Output from \Pyrrho{} (see Section~\ref{subsec:pyrrho}) enables us to do this already if we manually extract $\alpha$-values from each equation file produced and perform additional processing in, for example, spreadsheet software. This introduces a high potential for human error, however, and requires considerable additional data processing work that can be readily automated using the \Pacpal{} DSL (see Figure~\ref{fig:pacpal-skeptic-eg}). 




\begin{table}[ht] \tableFontSize
\caption{All 28 policies investigated in this work, ranked according to their $\alpha$-values given by \Skeptic{} in proportional reselection mode for each of the 3 datasets studied. Policy ranking performed by \Pacpal{}.}
\begin{tabular}{|l|l|l|l|l|}
\hline
Policy      & Yahoo & RockYou & LinkedIn & Average \\ \hline
3class16    & 1     & 1       & 2        & 1.33    \\ \hline
basic20     & 3     & 5       & 1        & 3       \\ \hline
2word16     & 2     & 4       & 5        & 3.67    \\ \hline
2class16    & 7     & 3       & 3        & 4.33    \\ \hline
3class12    & 4     & 2       & 8        & 4.67    \\ \hline
symbol10    & 9     & 8       & 9        & 8.67    \\ \hline
2word12     & 8     & 7       & 11       & 8.67    \\ \hline
symbol9     & 5     & 15      & 7        & 9       \\ \hline
2class12    & 15    & 6       & 12       & 11      \\ \hline
basic14     & 18    & 12      & 4        & 11.33   \\ \hline
comp8       & 6     & 9       & 19       & 11.33   \\ \hline
basic16     & 19    & 13      & 6        & 12.67   \\ \hline
upper9      & 11    & 10      & 18       & 13      \\ \hline
upper10     & 12    & 11      & 17       & 13.33   \\ \hline
basic12     & 20    & 14      & 10       & 14.67   \\ \hline
symbol8     & 14    & 18      & 14       & 15.33   \\ \hline
upper7      & 10    & 17      & 20       & 15.67   \\ \hline
symbol7     & 16    & 16      & 16       & 16      \\ \hline
digit10     & 17    & 20      & 13       & 16.67   \\ \hline
upper8      & 13    & 19      & 22       & 18      \\ \hline
basic10     & 21    & 21      & 15       & 19      \\ \hline
digit9      & 22    & 23      & 21       & 22      \\ \hline
digit7      & 24    & 22      & 24       & 23.33   \\ \hline
digit8      & 25    & 24      & 23       & 24      \\ \hline
basic9      & 23    & 26      & 25       & 24.67   \\ \hline
dictionary8 & 26    & 25      & 26       & 25.67   \\ \hline
basic7      & 27    & 28      & 27       & 27.33   \\ \hline
basic8      & 28    & 27      & 28       & 27.67   \\ \hline
\end{tabular}
\label{tbl:ideal-policy-table}
\end{table}

Rankings obtained using \Pacpal{} are shown in Table~\ref{tbl:ideal-policy-table}. Crucially, we do not require any access to the password data itself to produce these rankings, and thus we avoid the ethical issues involved in propagating user password data while retaining our ability to justify and reproduce these rankings as-needed. We make the \Pacpal{} scripts and equation files necessary to reproduce these results freely available\footnote{Access these here:
\url{https://github.com/sr-lab/skeptic-example-results}}.

\paragraph{\textbf{Findings}} We demonstrate that it is possible to use the \Skeptic{} toolchain to inform password composition policy choice, and that using the \Pacpal{} DSL this can be done without any additional manual data processing step.

\subsection{Policy Immunity} \label{subsec:pol-immunity}
In this section, we demonstrate the utility of encoding password composition policies in the \textit{Coq} proof assistant using \Authority{} (see Section~\ref{subsec:authority}) by formally verifying the immunity or vulnerability of 14 password composition policies to the password guessing attacks utilised by the \textit{Mirai} and \textit{Conficker} botnet worms. We achieve this by encoding the notion of vulnerability or immunity to concrete dictionaries of password guesses in Coq and devising a simple \texttt{simulate} tactic to prove, by dynamic simulation, assertions that a password composition policy either does or doesn't confer immunity to a guessing attack (see Figure~\ref{fig:immunity-proofs}).

\begin{figure}[ht] \codeSampleFontSize
    \centering
    \begin{minted}{coq}
(* The `basic14` policy is immune to Mirai. *)
Example basic14_mirai_immune :
  immune "basic14" mirai_dict.
Proof.
  simulate.
Qed.
    \end{minted}
    \caption{Examples of a proof in \textit{Coq}, showing that the policy named \texttt{basic14} renders a system immune to a guessing attack by the \textit{Mirai} malware.}
    \label{fig:immunity-proofs}
\end{figure}

\subsubsection{Mirai} \label{subsubsec:mirai}
\textit{Mirai} is a piece of malware that targets network-enabled devices running Linux, recruiting them into a botnet that has been used in several high-profile and extremely disruptive distributed denial-of-service (DDoS) attacks to date \cite{kolias2017ddos}. In order to propagate, \textit{Mirai} scans IP address ranges for devices with Telnet enabled. Upon locating a potentially vulnerable device, the malware will try a dictionary of 62 username/password combinations (containing 46 unique passwords) containing the factory defaults of a number of common internet-of-things (IoT) devices including CCTV cameras, home routers, and network-enabled printers \cite{antonakakis2017understanding}.

Using \textit{Coq}, at the level of the \Authority{}, we modelled the attack used by \textit{Mirai} to gain access to a device---a dictionary attack consisting of 46 specific guesses. From here, we were able to determine for a selection of the password composition policies by Shay et al.~\cite{shay2016designing} whether or not they render a device immune to \textit{Mirai} when enforced by prohibiting the creation of any vulnerable password. 

We are confident that these results (see Table~\ref{tbl:shay-mirai-immunity})
would be useful to any company producing Linux-based network-enabled devices. By configuring their devices with a password policy immune to compromise by \textit{Mirai} (such as basic16) before shipping, they are granted assurance that their product cannot be configured to become vulnerable.

\begin{table}[ht] \tableFontSize
    \centering
    \caption{Whether or not each password composition policy provides immunity to the dictionary attack used by the \textit{Mirai} worm, as verified from within Coq.}
    \label{tbl:shay-mirai-immunity}
    \begin{tabular}{|p{0.2\linewidth}|p{0.6\linewidth}|}
        \hline
        Immune & basic14, basic16, basic20, 2class16, 2word16, 3class16, comp8 \\
        \hline
        Vulnerable & basic7, basic8, basic9, basic12, 2class12, 2word12, 3class12 \\ 
        \hline
    \end{tabular}
\end{table}
   
\subsubsection{Conficker} \label{subsubsec:conficker}
Another botnet worm, \textit{Conficker} \cite{shin2012large}, which first emerged in 2008, remains a considerable threat even today through its use of several different propagation vectors to spread. One of these is a dictionary attack on password-protected administrative shares on Windows systems, which if successful allows the worm to write itself to disk on the remote machine and infect it.
The dictionary used by \textit{Conficker} for this purpose is, again, quite small containing only 182 passwords (including the empty password). By encoding our password composition policies from within Coq, we can ascertain whether each password policy from Shay et al.~\cite{shay2016designing} confers immunity against this attack as we did for Mirai. The results of this analysis are shown in Table~\ref{tbl:shay-conficker-immunity}.
   
\begin{table}[ht] \tableFontSize
    \centering
    \caption{Whether or not each password composition policy provides immunity to the dictionary attack used by the \textit{Conficker} worm, as verified from within Coq.}
    \label{tbl:shay-conficker-immunity}
    \begin{tabular}{|p{0.2\linewidth}|p{0.6\linewidth}|}
        \hline
        Immune & basic14, basic16, basic20, 2class12, 2class16, 2word12, 2word16, 3class12, 3class16, comp8 \\
        \hline
        Vulnerable & basic7, basic8, basic9, basic12 \\ 
        \hline
    \end{tabular}
\end{table}

Interestingly, if any of the policies analysed here are immune to \textit{Mirai}, they are also immune to \textit{Conficker} (i.e. the set of policies here that confer immunity to \textit{Mirai} are a subset of those that confer immunity to \textit{Conficker}). We anticipate that researchers will be able to use \Skeptic{} like this to discover policies immune to attack from a wide range of malware.

\section{Conclusion}\label{sec:conclusion}
In this work, we have demonstrated a new methodology for automatically, rigorously and justifiably selecting the most appropriate choice of password composition policy from a list of candidates. We achieve this by using a user behaviour model and password composition policy to induce a change in password probability distributions derived from large leaked password databases. We then take the uniformity of these distributions as a proxy for their security, demonstrating the validity of this approach by using it to closely reproduce results from two previous studies, one which collected passwords from users under specific password composition policies \cite{shay2016designing} and one which made use of large breached password datasets \cite{weir2010testing}. We find that our approach has the advantage of being attack-independent and broadly applicable, with its only assumption being that the attacker attempts to guess more common passwords first, but also that this comes at the expense of the ability to reason accurately about more attacks specifically tailored to target a particular system.

We have also described and presented \Skeptic{}, an implementation of this methodology as a software toolchain consisting of: \Authority{}, a metaprogramming utility for encoding policies in arbitrary representations; \Pyrrho{} a user behaviour model to redistribute probability according to these policies under different assumptions about user password reselection behaviour; and finally \Pacpal{}, a straightforward DSL to make the results of this process accessible to professionals working in the field. In addition, we have used this tool to obtain new results, including: a ranking of all 28 password composition policies studied in this work according to their expected effectiveness at mitigating password guessing attacks, under various assumptions about user password reselection behaviour; a demonstration that under some user behaviour models, certain password composition policies can have a negative effect on password security; and formal verification of the immunity of some password composition policies to the password guessing attacks employed by the \textit{Mirai} and \textit{Conficker} malware. 

\subsection{Future Work} \label{subsec:future-work}
We are excited about the future of this project, with the design of machine learning-based user behaviour models for password reselection representing a particularly promising potential future research direction. We also plan to expand the capabilities of \Pacpal{} to increase its utility, and explore the possibility of employing the power-law equations fitted by \Pyrrho{} in conjunction with existing password strength estimation algorithms to estimate the success probability of concrete password guessing attacks given as lists of strings.

We are also interested in devising tools and techniques to allow the synthesis of formally verified password composition policy enforcement software such as that by Ferreira et al.~\cite{ferreira2017certified} from models of password guessing attacks, informed by policy rankings produced by \Skeptic{}. Attack-defence trees in particular \cite{kordy2011foundations} appear promising as an intuitive formal representation of password guessing attacks and their mitigation measures from which password composition policies might be synthesised. We have taken some steps towards producing a user-friendly software interface for non-expert users to interact with \Skeptic{} and its satellite tooling with the \Passlab{} project \cite{johnson2020passlab}, and we believe with further implementation work we will be able to realise a fully-fledged graphical tool for defensive password security.

\paragraph{\textbf{Acknowledgements}} The authors would like to thank the anonymous reviewers,
whose expertise and insight was invaluable in shaping it.

\bibliographystyle{ACM-Reference-Format}
\bibliography{main}

\clearpage 

\appendix
\begin{landscape}
  \begin{table}
    \centering
    \caption{Appendix. A complete set of policy $\alpha$-values rankings for policies evaluated in \cite{shay2016designing} under each different macrobehaviour studied.}
    \resizebox{\columnwidth}{!}{
      \begin{tabular}[c]{|*{16}{l|}}
        \hline
        & & \multirow{2}{*}{Policy} & \multicolumn{4}{c|}{Yahoo} & \multicolumn{4}{c|}{RockYou} & \multicolumn{4}{c|}{LinkedIn}\\
      \cline{4-15}
      & & & Shay & Skeptic & $\alpha$ & Distance & Shay & Skeptic & $\alpha$ & Distance & Shay & Skeptic & $\alpha$ & Distance\\
      \hline
      \multirow{32}{*}{\rotatebox{90}{Reselection modes}} & \multirow{8}{*}{\rotatebox{90}{Null}} & 3class16 & 1 & 1 & -0.00015790845 & 0 & 1 & 1 & -0.00480967797 & 0 & 1 & 2 & -0.00511970014 & 1\\
      \cline{3-15}
      & & basic20 & 2 & 2 & -0.00017481256 & 0 & 2 & 2 & -0.00773612979 & 0 & 2 & 1 & -0.00206544273 & 1\\
      \cline{3-15}
      & & 2word16 & 3 & 3 & -0.00034446767 & 0 & 3 & 3 & -0.01310526071 & 0 & 3 & 3 & -0.01271757597 & 0\\
      \cline{3-15}
      & & basic16 & 4 & 6 & -0.01237917795 & 2 & 4 & 7 & -0.11203436164 & 3 & 4 & 4 & -0.11099256297 & 0\\
      \cline{3-15}
      & & 3class12 & 5 & 5 & -0.00946485322 & 0 & 5 & 5 & -0.01818160822 & 0 & 5 & 6 & -0.18384198515 & 1\\
      \cline{3-15}
      & & 2word12 & 6 & 7 & -0.01360245343 & 1 & 6 & 6 & -0.07942172914 & 0 & 6 & 5 & -0.17379245775 & 1\\
      \cline{3-15}
      & & comp8 & 7 & 4 & -0.00619759948 & 3 & 7 & 4 & -0.01573345733 & 3 & 7 & 7 & -0.21988288974 & 0\\
      \cline{3-15}
      & & basic12 & 8 & 8 & -0.16874098618 & 0 & 8 & 8 & -0.32090018785 & 0 & 8 & 8 & -0.44625701959 & 0\\
      \cline{2-15}
      & \multirow{8}{*}{\rotatebox{90}{Proportional}} & 3class16 & 1 & 1 & -0.15000000183 & 0 & 1 & 1 & -0.32803183792 & 0 & 1 & 2 & -0.45101422402 & 1\\
      \cline{3-15}
      & & basic20 & 2 & 3 & -0.22731830237 & 1 & 2 & 4 & -0.45407429983 & 2 & 2 & 1 & -0.45052415132 & 1\\
      \cline{3-15}
      & & 2word16 & 3 & 2 & -0.18899750304 & 1 & 3 & 3 & -0.4346028884 & 0 & 3 & 3 & -0.52489585375 & 0\\
      \cline{3-15}
      & & basic16 & 4 & 7 & -0.45303574889 & 3 & 4 & 7 & -0.579615909 & 3 & 4 & 4 & -0.57099747919 & 0\\
      \cline{3-15}
      & & 3class12 & 5 & 4 & -0.28309796453 & 1 & 5 & 2 & -0.33753384767 & 3 & 5 & 5 & -0.58017546055 & 0\\
      \cline{3-15}
      & & 2word12 & 6 & 6 & -0.31745131738 & 0 & 6 & 5 & -0.49108150848 & 1 & 6 & 7 & -0.61490864585 & 1\\
      \cline{3-15}
      & & comp8 & 7 & 5 & -0.2965234856 & 2 & 7 & 6 & -0.54963875987 & 1 & 7 & 8 & -0.65135140868 & 1\\
      \cline{3-15}
      & & basic12 & 8 & 8 & -0.47954187505 & 0 & 8 & 8 & -0.58639470743 & 0 & 8 & 6 & -0.59158613934 & 2\\
      \cline{2-15}
      & \multirow{8}{*}{\rotatebox{90}{Extraneous}} & 3class16 & 1 & 1 & -0.04210526403 & 0 & 1 & 1 & -0.1732211426 & 0 & 1 & 2 & -0.25848766731 & 1\\
      \cline{3-15}
      & & basic20 & 2 & 4 & -0.15048415667 & 2 & 2 & 3 & -0.2410656647 & 1 & 2 & 1 & -0.2478302857 & 1\\
      \cline{3-15}
      & & 2word16 & 3 & 2 & -0.05134151255 & 1 & 3 & 4 & -0.2463640901 & 1 & 3 & 3 & -0.29777195789 & 0\\
      \cline{3-15}
      & & basic16 & 4 & 6 & -0.17558403806 & 2 & 4 & 7 & -0.38191467167 & 3 & 4 & 4 & -0.40219971884 & 0\\
      \cline{3-15}
      & & 3class12 & 5 & 5 & -0.15869415661 & 0 & 5 & 2 & -0.22171184179 & 3 & 5 & 6 & -0.43333756896 & 1\\
      \cline{3-15}
      & & 2word12 & 6 & 7 & -0.18670016936 & 1 & 6 & 6 & -0.3512831245 & 0 & 6 & 5 & -0.42869639987 & 1\\
      \cline{3-15}
      & & comp8 & 7 & 3 & -0.15048415667 & 4 & 7 & 5 & -0.29031771829 & 2 & 7 & 7 & -0.4594561195 & 0\\
      \cline{3-15}
      & & basic12 & 8 & 8 & -0.35504148566 & 0 & 8 & 8 & -0.49858696195 & 0 & 8 & 8 & -0.53008440019 & 0\\
      \cline{2-15}
      & \multirow{8}{*}{\rotatebox{90}{Convergent}} & 3class16 & 1 & 7 & -1.33181526992 & 6 & 1 & 2 & -0.73706003039 & 1 & 1 & 5 & -0.84807451306 & 4\\
      \cline{3-15}
      & & basic20 & 2 & 8 & -1.65587234842 & 6 & 2 & 7 & -0.86310442053 & 5 & 2 & 8 & -0.90303209873 & 6\\
      \cline{3-15}
      & & 2word16 & 3 & 6 & -1.33177869336 & 3 & 3 & 5 & -0.79623023624 & 2 & 3 & 7 & -0.89905475536 & 4\\
      \cline{3-15}
      & & basic16 & 4 & 5 & -1.02369206677 & 1 & 4 & 6 & -0.85713632354 & 2 & 4 & 3 & -0.79663046993 & 1\\
      \cline{3-15}
      & & 3class12 & 5 & 2 & -0.77450139244 & 3 & 5 & 1 & -0.66271940447 & 4 & 5 & 2 & -0.77997709357 & 3\\
      \cline{3-15}
      & & 2word12 & 6 & 3 & -0.82018732762 & 3 & 6 & 3 & -0.74833314449 & 3 & 6 & 4 & -0.83475601093 & 2\\
      \cline{3-15}
      & & comp8 & 7 & 4 & -0.87004936668 & 3 & 7 & 8 & -0.92869922291 & 1 & 7 & 6 & -0.84854866977 & 1\\
      \cline{3-15}
      & & basic12 & 8 & 1 & -0.77238736541 & 7 & 8 & 4 & -0.77957401152 & 4 & 8 & 1 & -0.73119269609 & 7\\
      \hline
    \end{tabular}
  }
    \label{tbl:complete-results}
  \end{table}
  \end{landscape}

\end{document}